\documentclass[useAMS,a4,usenatbib]{mn2e}
\input psfig.sty

\usepackage{floatflt, epsfig}
\usepackage{epstopdf}

\def \chisq  {\ifmmode  \chi^2   \else  $\chi^2$  \fi}  
\def \spose#1{\hbox  to 0pt{#1\hss}}  
\def \lta{\mathrel{\spose{\lower 3pt\hbox{$\sim$}}\raise  2.0pt\hbox{$<$}}}
\def \gta{\mathrel{\spose{\lower  3pt\hbox{$\sim$}}\raise 2.0pt\hbox{$>$}}}

\def \kms {\ifmmode  \,\rm km\,s^{-1} \else $\,\rm km\,s^{-1}  $ \fi }
\def \kpc {\ifmmode  {\rm~kpc}  \else ${\rm~kpc}$\fi}  
\def \pc {\ifmmode  {\rm~pc}  \else ${\rm~pc}$ \fi  }  
\def \Gyr {\ifmmode  {\rm~Gyr}  \else ${\rm~Gyr}$\fi}
\def \Msun {\ifmmode M_{\odot} \else $M_{\odot}$ \fi} 
\def \Lsun {\ifmmode L_{\odot} \else $L_{\odot}$ \fi} 
\def \Rsun {\ifmmode R_{\odot} \else $R_{\odot}$ \fi} 
\def \Msunpyr {\ifmmode M_{\odot}{\rm~yr}^{-1} \else $M_{\odot}{\rm~yr}^{-1}$ \fi} 
\def \hMsun {\ifmmode h^{-1}\,\rm M_{\odot} \else $h^{-1}\,\rm M_{\odot}$ \fi}

\def \LCDM {\ifmmode \Lambda{\rm CDM} \else $\Lambda{\rm CDM}$ \fi}
\def \sig8 {\ifmmode \sigma_8 \else $\sigma_8$ \fi} 
\def \OmegaM {\ifmmode \Omega_{\rm M} \else $\Omega_{\rm M}$ \fi} 
\def \OmegaL {\ifmmode \Omega_{\rm \Lambda} \else $\Omega_{\rm \Lambda}$\fi} 
\def \Deltavir {\ifmmode \Delta_{\rm vir} \else $\Delta_{\rm vir}$ \fi}
\def \rhocrit {\ifmmode \rho_{\rm crit} \else $\rho_{\rm crit}$ \fi}
\def \rhou {\ifmmode \rho_{\rm u} \else $\rho_{\rm u}$ \fi}
\def \zc {\ifmmode z_{\rm c} \else $z_{\rm c}$ \fi}

\def \rhos {\ifmmode \rho_{\rm s} \else $\rho_{\rm s}$ \fi} 
\def \rs {\ifmmode r_{\rm s} \else $r_{\rm s}$ \fi} 
\def \cvir {\ifmmode c_{\rm vir} \else $c_{\rm vir}$ \fi} 
\def \Rvir {\ifmmode r_{\rm vir} \else $R_{\rm vir}$ \fi}
\def \Vvir {\ifmmode V_{\rm  vir} \else  $V_{\rm vir}$  \fi} 
\def \Mvir {\ifmmode M_{\rm  vir} \else $M_{\rm  vir}$ \fi}  
\def \Nvir {\ifmmode N_{\rm  vir} \else $N_{\rm  vir}$ \fi}  
\def \Jvir {\ifmmode J_{\rm vir} \else $J_{\rm vir}$ \fi} 
\def \Evir {\ifmmode E_{\rm vir} \else $E_{\rm vir}$ \fi} 
\def \vvir {\ifmmode v_{\rm vir} \else $v_{\rm vir}$ \fi} 
\def \lam {\ifmmode \lambda  \else $\lambda$ \fi} 
\def \lamp {\ifmmode \lambda^{\prime} \else $\lambda^{\prime}$  \fi} 
\def \Vmax {\ifmmode V_{\rm  max} \else  $V_{\rm max}$  \fi} 
\def \Mdm {\ifmmode M_{\rm  dm} \else $M_{\rm  dm}$ \fi}

\def \Mgas {\ifmmode M_{\rm gas} \else $M_{\rm gas}$ \fi} 
\def \Mcg {\ifmmode M_{\rm cg} \else $M_{\rm cg}$\fi} 
\def \Mhg {\ifmmode M_{\rm hg} \else $M_{\rm hg}$ \fi} 
\def \Mdisc {\ifmmode M_{\rm disc} \else $M_{\rm disc}$ \fi} 
\def \Md {\ifmmode M_{\rm d} \else $M_{\rm d}$ \fi} 
\def \Mda {\ifmmode M_{\rm d,0\%} \else $M_{\rm d,0\%}$ \fi} 
\def \Mdb {\ifmmode M_{\rm d,20\%} \else $M_{\rm d,20\%}$ \fi} 
\def \Mdc {\ifmmode M_{\rm d,40\%} \else $M_{\rm d,40\%}$ \fi} 
\def \md {\ifmmode m_{\rm d} \else $m_{\rm d}$ \fi} 
\def \Mb {\ifmmode M_{\rm b} \else $M_{\rm b}$ \fi} 
\def \Mbh {\ifmmode M_{\rm b,pri} \else $M_{\rm b,pri}$ \fi} 
\def \Mbs {\ifmmode M_{\rm b,sat} \else $M_{\rm b,sat}$ \fi} 
\def \zo {\ifmmode z_{0} \else $z_{0}$ \fi} 
\def \rd {\ifmmode r_{\rm d} \else $r_{\rm d}$ \fi}
\def \rg {\ifmmode r_{\rm g} \else $r_{\rm g}$ \fi}
\def \rb {\ifmmode r_{\rm b} \else $r_{\rm b}$\fi}
\def \rs {\ifmmode r_{\rm s} \else $r_{\rm s}$\fi}
\def \rc {\ifmmode r_{\rm c} \else $r_{\rm c}$\fi}
\def \rvir {\ifmmode r_{\rm vir} \else $r_{\rm vir}$\fi}
\def \rbh {\ifmmode r_{\rm b,pri} \else $r_{\rm b,pri}$ \fi} 
\def \rbs {\ifmmode r_{\rm b,sat} \else $r_{\rm b,sat}$ \fi}

\title[Hot gaseous halo in minor mergers] 
{The effects of a hot gaseous halo on disc thickening in galaxy minor mergers}

\author[B. P. Moster et al.] {Benjamin P. Moster$^{1,2}$
 \thanks{moster@mpa-garching.mpg.de}, Andrea V. Macci\`o$^{2}$, Rachel S. Somerville$^{3,4}$,
 \newauthor{Thorsten Naab$^{1}$, T. J. Cox$^{5}$}\\ 
  $^1$ Max-Planck Institut f\"ur Astrophysik, Karl-Schwarzschild Stra\ss e 1, 85748 Garching, Germany\\
  $^2$ Max-Planck-Institut f\"ur Astronomie, K\"onigstuhl 17, 69117 Heidelberg, Germany\\ 
  $^3$ Space Telescope Science Institute, Baltimore MD 21218\\
  $^4$ Department of Physics and Astronomy, Johns Hopkins University, Baltimore MD 21218\\
  $^5$ Carnegie Observatories, 813 Santa Barbara Street, Pasadena, CA 91101, USA\\
}

\begin{document} 
              
\date{\today}
              
\pagerange{\pageref{firstpage}--\pageref{lastpage}}\pubyear{2011} 
 
\maketitle 

\label{firstpage}
             
\begin{abstract}

We employ hydrodynamical simulations to study the effects of
dissipational gas physics on the vertical heating and thickening of
disc galaxies during minor mergers. For the first time we present a
suite of simulations that includes a diffuse, rotating, cooling, hot
gaseous halo, as predicted by cosmological hydrodynamical simulations
as well as models of galaxy formation.
We study the effect of this new gaseous component on the vertical
structure of a Milky Way-like stellar disc during 1:10 and 1:5
mergers.
For 1:10 mergers we find no increased final thin disc scale height compared to
the isolated simulation, leading to the conclusion that thin discs can be present
even after a 1:10 merger if a reasonable amount of hot gas is present.
The reason for this is the
accretion of new cold gas, leading to the formation
of a massive new thin stellar disc that dominates the surface
brightness profile.
In a previous study, in which we included only cold gas in the disk,
we showed that the presence of cold gas decreased the thickening by a
minor merger relative to the no-gas case. Here, we show that the
evolution of the scale height in the presences of a cooling hot halo
is dominated by the formation of the new stellar disc.
In this scenario, the thick disc is the old stellar disc that has been thickened in
a minor merger at $z\gta1$, while the thin disc is the new stellar disc
that reforms after this merger.
When galactic winds are also considered, the final scale height is
larger due to two effects. First, the winds reduce the star-formation
rate leading to a less massive new stellar disc, such that the
thickened old disc still dominates. Second, the winds exert a pressure
force on the gas in the disc, leading to a shallower gas profile and
thus to a thicker new stellar disc.
In addition, we study the evolution of the scale height during a 1:5 merger
and find that a thin disc can be present even after this merger, provided
enough hot gas is available.
The final scale height in our simulations depends on the mass of the
hot gaseous halo, the efficiency of the winds and the merger mass ratio. We
find post-merger values in the range $0.5\lta\zo\lta1.0\kpc$ in good
agreement with observational constraints by local galaxies.

\end{abstract}


\setcounter{footnote}{1}

\section{Introduction}
\label{sec:intro}

In the current paradigm of structure formation, large objects, such as
galaxies or clusters, are believed to form hierarchically, through a
'bottom-up' \citep{white1978} process of merging. In modern galaxy
formation theories, this merging process drives the evolution of many
galaxy properties. While major (near-equal mass) mergers can fully
transform disc-dominated systems into elliptical galaxies
\citep{toomre1977,negroponte1983,hernquist1992,naab2003,cox2006},
minor (unequal mass) mergers are believed to merely thicken the
galactic disc \citep{quinn1993,brook2004,bournaud2005,read2008,qu2011}.

About a decade ago, N-body simulations attained sufficient dynamic
range to reveal that, in Cold Dark Matter (CDM) models, all haloes
should contain a large number of embedded subhaloes that survive the
collapse and virialization of the parent structure
\citep{klypin1999,moore1999}.
Studies on merger statistics find that
roughly half of the mass delivery through mergers into
a dark matter halo of mass $M_h$ is due to systems with masses
$M_{\rm sat}=(0.03-0.3)M_h$. A large fraction (95 per cent)
of Milky Way (MW) sized haloes have accreted a satellite with a virial mass
comparable with the total mass of the MW disc ($\sim0.03M_h)$. About
70 per cent of the haloes experienced a 1:10 merger since $z\sim1$, while
$\sim40$ per cent had a 1:5 merger \citep{stewart2008,fakhouri2010,genel2010}.

This large population of merging
satellites has raised the question of whether mergers are {\em too}
common in the $\Lambda$CDM scenario. Numerous studies have
questioned whether thin, dynamically fragile discs such as the one
observed in the MW can survive this bombardment by incoming satellites
and found that the answer depends quite sensitively on the mass ratio of
the merging galaxies \citep[e.g.][]{quinn1986,toth1992,quinn1993,
walker1996,sellwood1997,velazquez1999,font2001,benson2004,
gauthier2006,kazantzidis2008,kazantzidis2009,villalobos2008,
purcell2009}. There seems to be a consensus that the
main danger to thin discs is from events with a dark matter mass ratio of
($\sim$1:10). The ubiquitous mergers with a lower mass satellite of
$M_{\rm sat}\lta M_{\rm disc}$ do not destroy the thin disc \citep{kazantzidis2008}.
On the other hand, mergers with a very massive satellite are rare such
that only few disc systems are affected. However, mergers with an intermediate mass
ratio of 1:10 are frequent enough for MW-sized systems, and in dissipationless
simulations these events destroy the thin stellar disc. This would imply that thin
discs in MW-sized systems are rare, in disagreement with the observation
of many systems with thin discs in the Universe \citep[e.g.][]{schwarzkopf2000,yoachim2006}.

Previous studies of disc stability against satellite infall have often been affected by
numerical limitations or by analytic assumptions that have later been found to be
contradicting the $\Lambda$CDM model. Typical problems include initial discs
that are rather thick compared to observations of local disc galaxies \citep{quinn1993,
velazquez1999,font2001,villalobos2008}, the modelling of structural components as
rigid potentials \citep{quinn1986,quinn1993,sellwood1997}, the negligence of the dark
matter component of the satellite \citep{quinn1993,walker1996,velazquez1999}, and
the use of satellite orbital parameters that are in conflict with the standard cosmology.
Also in analytic models, simplifications had to be assumed, such as the local deposition
of the orbital energy of the satellite \citep{toth1992}, and the absence of global heating modes
\citep{benson2004}. More importantly, most studies so far, have neglected the gaseous
component in the galaxy.

However, the presence of a dissipative gas component is now known to
play an important role in galaxy mergers and is crucial in order to
reproduce basic properties of observed galaxies \citep{mihos1994,naab2006}.
Unlike stars and dark
matter, gas is able to cool radiatively, thereby losing kinetic
energy efficiently. Furthermore, as the gas is eventually transformed
into stars, a new rotating, kinematically cold stellar disc can form
during and after a merger. Therefore, a dissipative component in the
galactic disc has a stabilising effect.  \citet[][M10]{moster2010b}
found that when the presence of gas in the disc is taken into account,
the thickening of the stellar disc is reduced. For an initial disc gas
fraction of 20 (40) per cent, the thickening was reduced by 25 (50)
per cent with respect to collisionless simulations. Final disc scale
heights found in the simulations are in good agreement with studies of
the vertical structure of spiral galaxies.

In M10, we argued that the presence of gas can reduce disc heating via
two mechanisms: absorption of kinetic impact energy by the gas and/or
formation of a new thin stellar disc that can cause heated stars to
recontract towards the disc plane. As in the initial conditions only
the cold gas in the disc was accounted for (neglecting the gaseous
halo), most of the gas was consumed during the merger, and therefore
the regrowth of a new thin disc had a negligible impact on the scale
height of the post-merger galaxy. This led to the conclusion that the
main process that suppresses disc thickening in the absence of a
continuous supply of fresh gas is the absorption of impact energy by
the gas. Furthermore, M10 concluded that in order to reform a new thin
disc comparable in mass to the old disc, an external fueling
reservoir, such as cooling and accretion from a gaseous halo, is
needed.

While M10 focused their attention only on the cold gas component
within the disc, semi-analytic models of galaxy formation
\citep{kauffmann1993,bower2006,somerville2008a} as well as full
cosmological hydrodynamic simulations \citep[e.g.][]{toft2002,
  sommerlarsen2006,johansson2009b,rasmussen2009,stinson2010,hansen2010}
both predict a large amount of hot gas in quasi-hydrostatic
equilibrium within the gravitational potential of the dark matter
halo. Cooling and accretion of gas from these haloes can grow the
discs of spiral galaxies
\citep[e.g.][]{abadi2003,sommerlarsen2003}. It is therefore expected
that the presence of a hot gaseous halo leads to the growth of a new
thin stellar disc that may affect the evolution of the disc scale
height.

In this paper we expand our previous work on disk stability by
studying the effects of gas cooling from a hot reservoir on the
vertical structure of the disc in a MW-like galaxy during minor
mergers. We extend our previous study by re-simulating the fiducial
intitial conditions from M10 (1:10 (dark matter mass ratio) merger
with 20 per cent of cold gas in the disc, an orbital inclination of
$60^{\circ}$ and an initial disc scale height of $\zo = 0.4\kpc$), now
including a hot gaseous halo. We assume that the mass of hot gas in
the halo is such that the whole system contains 85 percent of the
universal baryon fraction.
We also run the same system in isolation, in order to study how the
accretion of gas from the halo acts on the scale height.

We then present a detailed analysis of the effects of this new gas
component. We focus our attention on the transformation of the initial
stellar disc into a thick disc and the regrowth of a new thin stellar
disc after the mergers. The continuous accretion of gas and the
reformation of a massive thin disc is expected to affect the evolution
of the disc scale height. On the one hand the potential of the new
disc can cause heated stars to recontract towards the disc plane; on
the other hand the new thin disc can dominate the surface brightness
and therefore lead to a lower observed scale height.

We further study the effects of galactic winds on the evolution of the
scale height. As winds reduce the SFR, the new stellar disc is
expected to be less massive such that the surface brightness profile
is dominated by the thickened old disc. In addition, winds can exert a
pressure force on the gas in the disc, leading to a thicker gas disc
and thus to a thicker new disc than in the windless case. We
demonstrate how these two effects impact the evolution of the scale
height in our simulations.

Finally, we investigate how the same galactic disc evolves in a 1:5
merger (dark matter mass ratio). As the effects from the hot gaseous
halo are expected to further decrease the final scale height, we study
whether the disc can survive such a merger if the hot component is
included and whether the final scale height is in agreement with
observational constraints.

The paper is organised as follows: in section \ref{sec:nsim} we
provide a brief summary of the {\sc GADGET-2} code and the initial
conditions.  We also summarise how the initial conditions have been
expanded in order to include a rotating hot gaseous halo. In section
\ref{sec:res10} we present our results for the 1:10 merger
simulations, focusing on the differences between the simulations with
and without a hot gaseous component and on the effects of the galactic
winds. In section \ref{sec:res5} we show the results for the 1:5 minor
merger simulations and study how the disc thickening is changed when a
hot halo is included. Finally, in section \ref{sec:conc} we summarise
and discuss our results and compare them to previous studies that have
neglected the gaseous halo.

\section{Numerical Simulations} 
\label{sec:nsim}

\begin{table*}
 \centering
 \begin{minipage}{140mm}
  \caption{Parameters kept constant for all simulations. Masses are in units of  $10^{10}\Msun$, scale and 
softening lengths are in units of \kpc~and pc, respectively}
  \begin{tabular}{@{}lrrrrrrrrrrrrr@{}}
  \hline
  System & \Mdm & \Mdisc & \Mb & \rd & \rg & \rb & \zo & $c$ & $N_{\rm dm}$ & $N_{\rm disc}$ & $N_{\rm bulge}$\\
 \hline
 \hline
Primary & 100 & 2.4 & 0.600 & 3.00 & 3.0 & 0.50 & 0.4 & 9.65 & 2 000 000 & 500 000 & 250 000\\
Sat (1:10) & 20 & 0.0 & 0.063 & 0.00 & 0.0 & 0.30 & 0.0 & 11.98 & 450 000 & 0 & 50 000\\
Sat (1:5) & 10 & 0.0 & 0.162 & 0.00 & 0.0 & 0.35 & 0.0 & 11.45 & 900 000 & 0 & 130 000\\
\hline
\label{t:conpar}
\end{tabular}
\end{minipage}
\end{table*}

\subsection{Numerical Code} 
\label{sec:ncode}

We employ the parallel TreeSPH-code {\sc GADGET-2}
\citep{springel2005a} to conduct the numerical simulations
presented in this paper. The gaseous component is evolved using the
Lagrangian Smoothed Particle Hydrodynamics
\citep[SPH][]{lucy1977,gingold1977,monaghan1992} technique, employed
in a formulation that conserves both energy and entropy
\citep{springel2002}. Radiative cooling is implemented for a
primordial mixture of hydrogen and helium following \citet{katz1996},
including a spatially uniform time- independent local photo-ionizing UV
background in the optically thin limit \citep{Haardt1996}.

The implementation of star formation and the associated heating by
supernovae (SN) follows the sub-resolution multiphase ISM model
developed by \citet{springel2003}. This model assumes that a thermal
instability operates above a critical density threshold $\rho_{th}$,
such that the ISM is a two-phase medium with cold clouds embedded in a
tenuous gas at pressure equilibrium. Stars form in dense regions on a
timescale chosen to match observations \citep{kennicutt1998} and
short-lived stars supply energy to the surrounding gas as supernovae
heating the diffuse phase.

The threshold density $\rho_{th}$ is determined self-consistently by
demanding that the equation of state (EOS) is continuous at the onset
of star formation. SN-driven galactic winds as proposed by
\citet{springel2003} are included in a subset of simulations. In this
model the mass-loss rate carried by the wind is proportional to the
star formation rate (SFR) $\dot M_w= \eta \dot M_*$, where the wind
efficiency is quantified by the mass-loading-factor $\eta$. The wind
is assumed to carry a fixed fraction of the supernova energy, leading
to a constant initial wind speed $v_w$.

We adopt the standard parameters for the multiphase model in order to
match the Kennicutt Law as specified in \citet{springel2003}. The star
formation timescale is set to $t_*^0 = 2.1\Gyr$, the cloud evaporation
parameter to $A_0 = 1000$ and the SN ``temperature'' to $T_{\rm
  SN}=10^8{\rm~K}$.  A \citet{salpeter1955} initial mass function
(IMF) is assumed, setting the mass fraction of massive stars to
$\beta=0.1$. We further adopt a mass-loading factor of $\eta = 2$ and
a wind speed of $v_w \sim 480 \kms$, typical for a MW-like galaxy at
low redshift. We do not include feedback from accreting black holes
(AGN feedback) in our simulations.

\subsection{Galaxy Models} 
\label{sec:nmod}

The galaxy models used in our simulations are constructed with the
method developed by \citet{springel2005b} with the extension to
include a hot gas halo as described by \citet[][M11]{moster2011a}. Each
primary system (i.e. the central galaxy) is composed of a cold gaseous
disc, a stellar disc and a stellar bulge with masses \Mcg, \Mdisc and
\Mb embedded in a halo that consists of hot gas and dark matter with
masses \Mhg and \Mdm. The satellite systems only consist of dark
matter and a spherical stellar component.

The dark matter halo has a \citet{hernquist1990} profile with a scale
radius $a$ corresponding to a Navarro-Frenk-White
\citep[NFW;][]{navarro1997} halo with a scale length \rs~and a
concentration parameter $c=\rvir/\rs$. The gaseous and stellar discs
are rotationally supported and have exponential surface density
profiles with an equal scale length $r_{\rm d}$. 
The vertical structure of the stellar disc is described by a radially
independent sech$^2$ profile with a scale height $z_0$, such that
the stellar density is described by
\begin{equation}
\rho_*(R,z)=\frac{M_{\rm disc}}{4\pi z_0 r_{\rm d}^2} \; {\rm sech}^2
\left( \frac{z}{z_0}\right) \; \exp \left( -\frac{R}{r_{\rm d}}\right) \; .
\end{equation}
The vertical velocity dispersion is set equal to the radial velocity
dispersion. The gas temperature is fixed by the EOS, rather than the
velocity dispersion. The vertical structure of the gaseous disc is
computed self-consistently as a function of the surface density by
requiring a balance of the galactic potential and the pressure given
by the EOS. The spherical stellar bulge is non-rotating and has a
\citet{hernquist1990} profile with a scale length \rb.

We model the hot gaseous component as a slowly rotating halo with a
spherical density profile \citep{moster2011a}.  The density
distribution follows the observationally motivated $\beta$-profile
\citep{cavaliere1976,jones1984,eke1998}:
\begin{equation}
\rho_{\rm hg}(r) = \rho_0 \left[1+\left(\frac{r}{\rc}\right)^2\right]^{-\frac{3}{2}\beta}\;.
\end{equation}
which has three free parameters: the central density $\rho_0$, the
core radius \rc~and the outer slope parameter $\beta$.

The temperature profile is fixed by assuming an isotropic model and
hydrostatic equilibrium inside the galactic potential. Furthermore,
the hot gaseous halo is rotating around the spin axis of the disc
with a specific angular momentum $j_{\rm hg}$ that is a multiple of
the specific angular momentum of the dark matter halo $j_{\rm dm}$
such that $j_{\rm hg}=\alpha j_{\rm dm}$. The angular momentum
distribution scales with the product of the distance from the spin
axis $R$ and the circular velocity at this distance: $j(R) \propto R
\; v_{\rm circ}(R)$.

\subsection{Simulation Parameters} 
\label{sec:npar}

\begin{table*}
 \centering
 \begin{minipage}{140mm}
  \caption{Parameters for the different simulation runs. Masses are in
    units of $10^{10}\Msun$, and softening lengths are in units of pc.}
  \begin{tabular}{@{}lrrrrrrrrrr@{}}
  \hline
  Run & $f_{\rm gas}$ &  \Mcg & \Mhg & $N_{\rm cg}$ & $N_{\rm hg}$ & $\lambda$ & $\eta$ & $v_{\rm wind}$ & $\theta$ & ratio\\
 \hline
 \hline
IA & 0.0 & 0.0 & 0.0 & 0 & 0 & 0.033 & 0.0 & 0 & - & -\\
IB & 0.2 & 0.6 & 0.0 & 31 250 & 0 & 0.034 & 0.0 & 0 & - & -\\
IBh & 0.2 & 0.6 & 11.5 & 31 250 & 650 000 & 0.034 & 0.0 & 0 & - & -\\
IBw & 0.2 & 0.6 & 0.0 & 31 250 & 0 & 0.034 & 1.0 & 480 & - & -\\
IBhw & 0.2 & 0.6 & 11.5 & 31 250 & 650 000 & 0.034 & 1.0 & 480 & - & -\\
\hline
MA10 & 0.0 & 0.0 & 0.0 & 0 & 0 & 0.033 & 0.0 & 0 & $60^{\circ}$ & 1:10\\
MB10 & 0.2 & 0.6 & 0.0 & 31 250 & 0 & 0.034 & 0.0 & 0 & $60^{\circ}$ & 1:10\\
MB10h & 0.2 & 0.6 & 11.5 & 31 250 & 650 000 & 0.034 & 0.0 & 0 & $60^{\circ}$ & 1:10\\
MB10w & 0.2 & 0.6 & 0.0 & 31 250 & 0 & 0.034 & 1.0 & 480 & $60^{\circ}$ & 1:10\\
MB10hw & 0.2 & 0.6 & 11.5 & 31 250 & 650 000 & 0.034 & 1.0 & 480 & $60^{\circ}$ & 1:10\\
\hline
MA5 & 0.0 & 0.0 & 0.0 & 0 & 0 & 0.033 & 0.0 & 0 & $60^{\circ}$ & 1:5\\
MB5 & 0.2 & 0.6 & 0.0 & 31 250 & 0 & 0.034 & 0.0 & 0 & $60^{\circ}$ & 1:5\\
MB5h & 0.2 & 0.6 & 11.5 & 31 250 & 650 000 & 0.034 & 0.0 & 0 & $60^{\circ}$ & 1:5\\
MB5w & 0.2 & 0.6 & 0.0 & 31 250 & 0 & 0.034 & 1.0 & 480 & $60^{\circ}$ & 1:5\\
MB5hw & 0.2 & 0.6 & 11.5 & 31 250 & 650 000 & 0.034 & 1.0 & 480 & $60^{\circ}$ & 1:5\\
\hline
\label{t:varpar}
\end{tabular}
\end{minipage}
\end{table*}

We select the same parameters for the fiducial galaxy models as M10.
For the primary disc galaxy system we use a virial mass of $M_{\rm dm}
= 10^{12}\Msun$ and a concentration parameter of $c=9.65$, following
\citet{maccio2008}. The stellar mass of the galaxy is determined using
the empirical relation derived by \citet{moster2010a}, which yields
$M_{\rm *,pri}=3\times10^{10}\Msun$, resulting in a stellar-to-halo mass
ratio of 1:33. Distributing 80 per cent of this
stellar mass into the disc results in a stellar disc mass of
$\Mdisc=2.4\times10^{10}\Msun$ and a bulge mass of
$\Mdisc=0.6\times10^{10}\Msun$.

For models with a cold gaseous disc we add
$\Mcg=0.6\times10^{10}\Msun$ such that the gas fraction in the disc is
20 per cent.  We set the scale lengths of the stellar and the gaseous
disc to $\rd=3\kpc$, which fixes the spin parameter to $\lambda=0.033$
for the case without gas and $\lambda=0.034$ for the models with cold
gas in the disc. We set the scale height of the stellar disc to
$\zo=0.4\kpc$ and the bulge radius to $\rb=0.5\kpc$.

In addition to the models employed in M10 we create a system that also
includes a hot gaseous halo. We fix the mass of this halo by assuming
that the baryonic fraction within the virial radius of the system is
85 per cent of the Universal one. This leads to a hot gas halo mass of
$\Mhg = 1.1\times10^{11}\Msun$ within \rvir.  For the density profile,
we adopt $\beta = 2/3$ \citep{jones1984} and $\rc=0.22\rs$
\citep{makino1998}. The spin parameter $\alpha$ has been constrained
by \citet{moster2011a} using isolated simulations of a MW-like
galaxy and demanding that the observed evolution of the average
stellar mass and scalelength be reproduced. M11 found the best
agreement with the observational constraints with a spin factor of
$\alpha=4$ which will also be assumed throughout this work.
In this way our disc galaxy models have a consistent evolution since $z=1$.

The satellite systems consist of only dark matter and a stellar
bulge. We employ two (dark matter) mass ratios for the merger, 1:10
and 1:5. For the 1:10 merger the dark matter mass of the satellite is
$M_{\rm dm} = 10^{11} \Msun$ with a concentration of $c=11.98$. The
dark matter mass of the satellite for the 1:5 merger is $M_{\rm dm} =
2\times10^{11} \Msun$ with a concentration of $c=11.45$. Using the
stellar-to-halo mass relation we derive a stellar mass of
$\Mb=6.3\times 10^{8}\Msun$ for the 1:10 merger and a stellar mass of
$\Mb=1.62\times 10^{9}\Msun$ for the 1:5 merger. The stellar-to-halo
mass ratio of the satellites is 1:159 and 1:62 for the 1:10 and 1:5 mergers,
respectively. Finally we fix the
bulge scale lengths at $\rb=0.3\kpc$ and $\rb=0.35\kpc$ for the 1:10
and 1:5 mergers, respectively.

The primary system is modelled with $N_{\rm dm}=2\times10^6$ dark
matter, $N_{\rm disc}=5\times10^5$ stellar disc, $N_{\rm
  cg}=3.1\times10^4$ gaseous disc, $N_{\rm bulge}=2.5\times10^5$ bulge
and $N_{\rm hg}=6.5\times10^5$ gaseous halo particles. The satellite system
has  $N_{\rm dm}=4.5\times10^5$ ($N_{\rm dm}=9\times10^5$) dark
matter and $N_{\rm bulge}=5\times10^4$ ($N_{\rm bulge}=1.3\times10^5$)
bulge particles for the 1:10 (1:5) merger.\footnote{This choice of particle
  numbers results in a resolution that
  is a factor of two lower in dark matter and initial stars and a
  factor of four lower in gas and newly formed stars than that
  employed by M10. The impact of this lowered resolution on our
  results is discussed in section \ref{sec:reso}.}

We set the gravitational softening length to $\epsilon = 70\pc,
100\pc$ and $140\pc$ for stellar, gas and dark matter particles,
respectively. The merger orbit has been chosen to be consistent with
cosmological simulations and has an eccentricity of $\epsilon=0.89$, a
pericentric distance of $r_{\rm min}=18\kpc$ and an initial separation
of $d_{\rm start}=120\kpc$. We use a prograde orbit with an angle of
$\theta=60^{\circ}$ between the spin axes of the disc and the orbit
and evolve the simulations for 6\Gyr.
Prograde orbits are expected to be more destructive to the disc than retrograde
mergers \citep{velazquez1999}.

We summarise the parameters that are kept constant for all simulations
in Table \ref{t:conpar}, and specify all parameters that differ for
the various simulation runs in Table \ref{t:varpar}. The different
simulations are labeled with the first letter I for isolated runs and
an M for mergers. The second letter specifies the gas fraction of the
disc (A for $f_{\rm gas}=0$ and B for $f_{\rm gas}=0.2$) followed by a
number that indicates the merger mass ratio. If a hot gaseous halo
included an \lq h\rq~is added and if galactic winds are included we
add a \lq w\rq.

\section{Results for the 1:10 merger}
\label{sec:min}
\label{sec:res10}

In order to study the evolution of the disc thickness, we compute the
edge-on projected surface density as a function of the distance to the galactic plane
at a distance of $\Rsun\sim8\kpc$ from the disc spin axis. We
determine the disc scale height $z_0$ by fitting a sech$^2$ function
to the edge-on surface density profile $\Sigma(z)$ at \Rsun
to be able to compare the results to observations.  We note that
values for $z_0$ are potentially larger than what one would obtain by
fitting to the 3d density profile, as the disc can develop
considerable flaring (i.e. the disc becomes thicker with increasing
distance from the spin axis).

\subsection{Evolution of the scale height and effects of cold gas}

Before studying the effects of mergers, it is important to first
address the stability of the initial disc model in isolation. This
allows us to disentangle instabilities of the isolated disc (both
physical and numerical) from the thickening that is caused by
mergers. The evolution of the disc scale height for the isolated
galaxy and 1:10 merger simulations for the case without galactic winds
is shown in Figure \ref{fig:z10halo}.

The result for the simulation of the isolated galaxy without gas (IA)
is given by the red dotted line. The scale height increases from an
initial value of $z_0=0.4\kpc$ to a final value of $z_0=0.5\kpc$,
which means that the instabilities of the isolated disc lead to an
increase of 25 per cent. This effect is mainly numerical and will be
further discussed in section \ref{sec:reso}. On the other hand, any
further increase of the scale height should be due to accretion
events. If cold gas is included in the disc, the final scale height is
slightly lower, due to the new thin stellar disc that forms from the
cold gaseous disc through star formation. Although this new disc is
thinner than the old disc, its mass is much lower, such that the total
scale height is very close to the result obtained for the dissipationless
case.

For the 1:10 merger simulations we find that the disc is considerably
thickened beyond the degree found for the isolated runs.  The scale
height for the merger without gas has a final value of $z_0=0.8\kpc$
which implies a thickening by a factor of 2 (red solid line in Figure
\ref{fig:z10halo}. In the simulation that includes an initial 20 per
cent cold gas in the disc (blue solid line) this thickening is reduced,
resulting in a final scale height of $z_0=0.7\kpc$. It has been shown
by M10 that the physical process that is causing this change in
behaviour is the absorption of kinetic impact energy of the satellite
by the gas component. The gas can efficiently radiate this energy away
by cooling. On the other hand, the reformation of a new thin disc has
little effect on the evolution of the scale height, as the final mass
of this new stellar disc ($M_{\rm *,new} \sim0.5\times10^{10}\Msun$)
is much lower then the initial old disc ($M_{\rm
  *,old}\sim2.4\times10^{10}\Msun$). As a result, the disc is
thickened during the first two encounters, while afterwards, the scale
height is roughly constant.

We note that the evolution of the scale height for the 1:10 merger
simulations is in very good agreement with the results presented by
M10, although the spatial resolution of the simulations analyzed here is lower
by a factor of two. See section \ref{sec:reso} for a discussion.

\subsection{Effects of the hot gaseous halo}

\begin{figure}
\psfig{figure=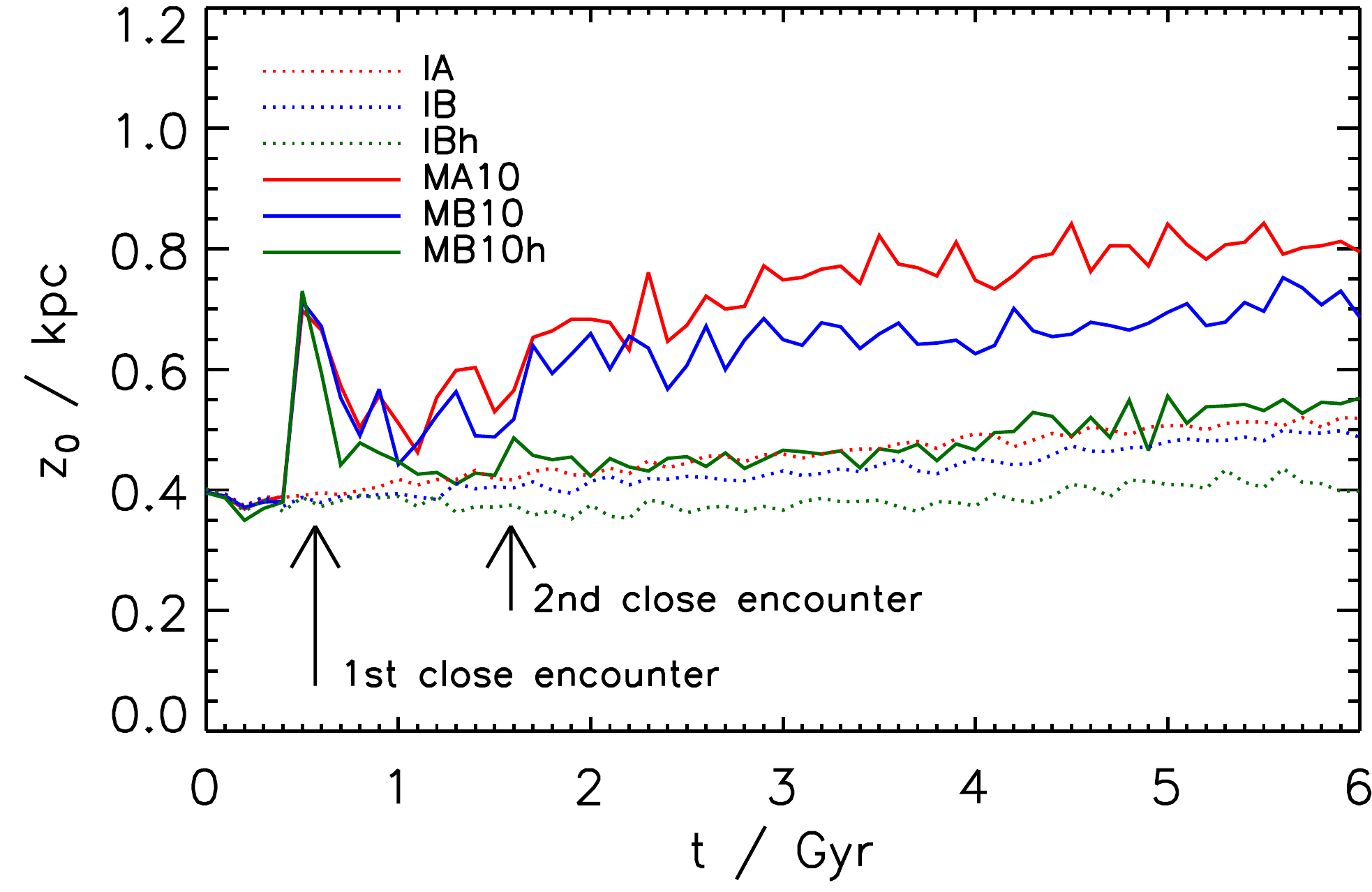,width=0.47\textwidth}
\caption{Evolution of the disc scale height for isolated and 1:10
  merger simulations with and without a gaseous halo for the case
  without galactic winds.  }
\label{fig:z10halo}
\end{figure}

All previous studies on disc heating and thickening have either
completely neglected the dissipational component, or have only
included the cold gas in the disc.  However, M10 concluded that the
reformation of a new thin stellar disc could reduce the total scale
height, if an external fuelling reservoir were available that could
provide enough material such that the new disc could obtain a mass
that is comparable to the old one. In this section, we study how the
scale height of the disc evolves if this reservoir is provided by a
rotating, cooling hot gaseous halo, as modelled by M11.

The resulting time evolution of \zo for the isolated run including a
hot gaseous component (IBh) is presented in Figure \ref{fig:z10halo}
(green dotted line). Although we have seen before that the initial stellar
disc does thicken in an isolated run, the scale height of the total
stellar disc is constant at $z_0=0.4\kpc$. The reason for this is the
formation of a new very thin stellar disc which has a mass that is even
larger than the old disc. As a result, this new disc dominates the
surface density profile and compensates for the numerical thickening
of the old disc. The scale height of the disc in the IBh simulation is
thus mostly determined by how the gas accretes from the halo, i.e. on
the profile of the cold gas disc.

In the 1:10 merger simulation including a hot gaseous halo (MBh, green solid
line), we find a completely different behaviour than in the run without a hot
component. While the scale height increases during the first and
second encounters, it decreases again after the encounters.  At the
end of the simulation, the scale height of all stars is equal to that in the
isolated run that does not include gas at all (and where all
thickening is due to numerical effects) with a final value of
$z_0=0.5\kpc$. This means that due to the gas accretion from the hot
halo and the resulting new thin stellar disc the overall scale height
shows no additional thickening due to the 1:10 merger with respect to
the isolated dissipationless simulation. However, the scale height is
still larger than in the isolated run including a hot gaseous halo.

\begin{figure}
\psfig{figure=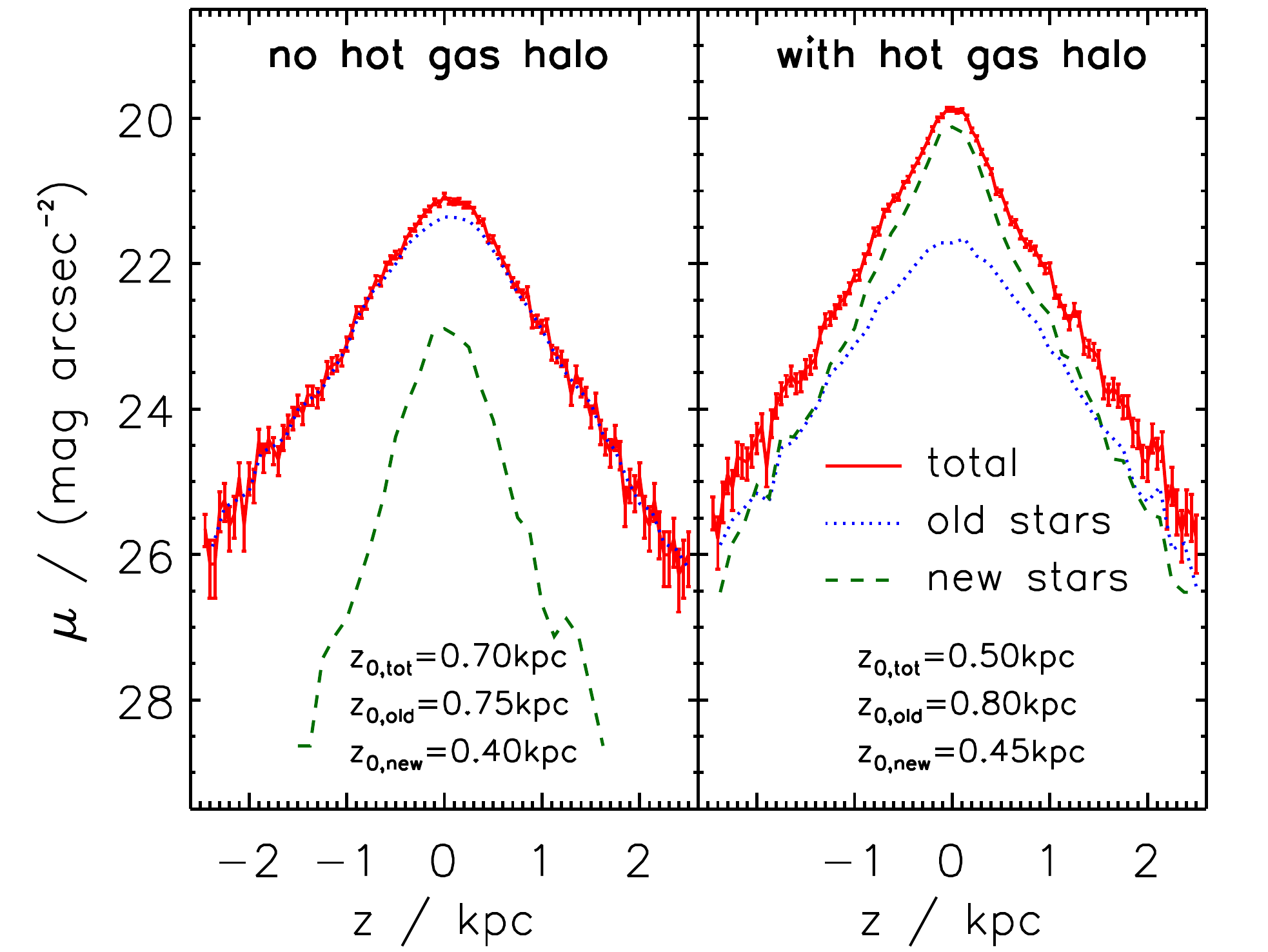,width=0.47\textwidth}
\caption{Final edge-on surface brightness profiles for the 1:10 merger
  simulations without winds at a projected radius of $8\kpc$ for old,
  new, and all stars. The left panel shows the run without a hot
  gaseous halo and the right panel gives the results for the run
  including the hot component (right panel). A mass-to-light ratio of
  $M/L=3(M/L)_{\odot}$ has been assumed, appropriate for the Milky Way
  in the $B$ band.}
\label{fig:cprofh}
\end{figure}

There are two possible effects that can cause this: one option is that
the potential of the new stellar disc forces heated stars to contract
again onto the disc plane; the other possibility is that the new disc
is both thin and massive, such that the contribution to the density
profile of the total disc is large. In order to investigate this
further, we compute the scale height at the end of the simulations for
old stars (i.e. stars which have been set up as initial conditions)
and new stars (i.e. stars that formed during the simulation). We
present the resulting surface brightness profiles for old, new, and
all stars in Figure \ref{fig:cprofh}, where we have assumed a
mass-to-light ratio of $M/L=3(M/L)_{\odot}$, appropriate for the MW in
the $B$ band \citep{zibetti2009}. We find a value for the scale height
of the new thin disc in MBh of $z_{0,new} = 0.45\kpc$ while the old
stellar disc has $z_{0,old} = 0.8\kpc$. Thus the reformation of a new
massive disc has not caused the old stars to contract onto the disc
plane. In the MB run, the scale height of the new disc is $z_{0,new} =
0.4\kpc$ and that of the old disc is $z_{0,old} = 0.75\kpc$. Thus,
although for MBh both the new and the old discs are thicker than the
respective ones of MB, the total scale height is smaller. This can
only be explained by the differences in mass of the new disc between
MB and MBh.

\begin{figure*}
\psfig{figure=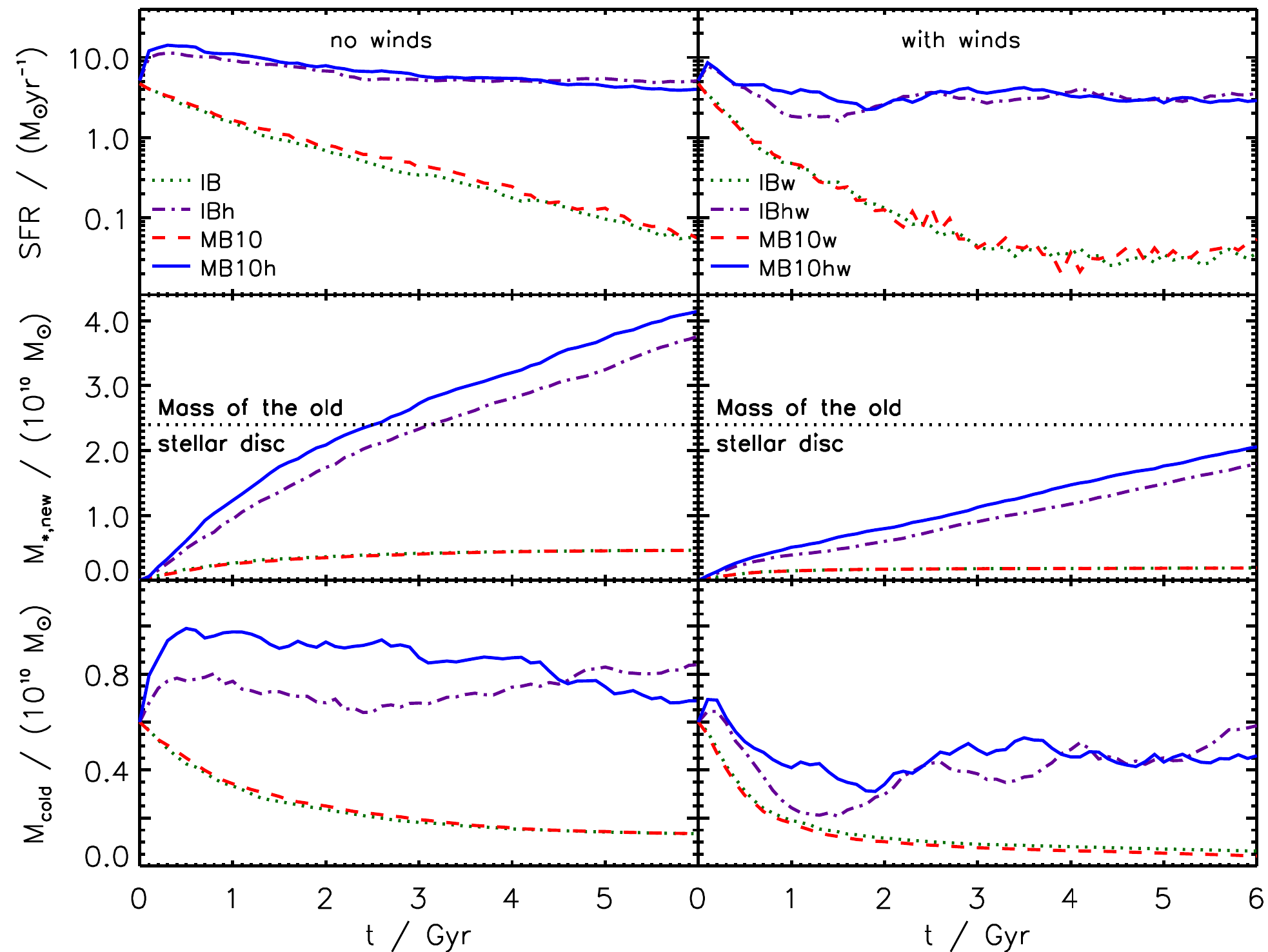,width=0.9\textwidth}
\caption{The rows from top to bottom show the SFR, the total mass of
  new stars formed during the simulation and the cold gas mass for the
  simulations with and without a gaseous halo. The panels on the left
  hand side give the results for the simulations without galactic
  winds, while the simulations with winds are presented on the right
  hand side.}
\label{fig:sfrminor}
\end{figure*}

To demonstrate this, we plot the SFR, the total mass of new stars
formed during the simulation and the cold gas mass in Figure
\ref{fig:sfrminor} (left panels). While the SFR after the second close
encounter is already very low for the MB simulation ($<1\Msunpyr$),
the MBh run maintains a high SFR throughout the simulation
($\sim5-10\Msunpyr$ for a Salpeter IMF). These SFRs lead to a final stellar
mass (after $6\Gyr$) that is roughly twice the initial value, consistent with results for
low redshift galaxies from subhalo abundance matching \citep{moster2010a} and semi-analytic
models of galaxy formation \citep{somerville2008a}. As a result of this the mass
of the new thin disc is much higher in the MBh run and even exceeds
that of the old stellar disc component at the end of the simulation.
As the new thin disc is more massive than the old disc, the
contribution to the total density profile and thus to the total scale
height is dominated by the new stellar disc. As a result the scale
height of the total disc in MBh is smaller than that of MB, although
both the new and the old discs are thicker than those found in
MB. This can also be seen in Figure \ref{fig:cprofh}: although the new
stars in MBh form a thicker disc than in MB, they dominate the total
surface brightness profile, such that it is thinner than the total
profile of MB.

Another effect we notice is
the formation of a more massive cold gaseous disc than the one that we started
with. The mass of the cold gaseous disc quickly rises from $M_{\rm
  cg}=0.6\times10^{10}\Msun$ to $M_{\rm cg}\sim10^{10}\Msun$ and only
decreases slightly through the course of the simulation. This means
that the disc gas fraction during the first two encounters of this
simulation is $\sim20$ percent (while the simulation without a hot
gaseous halo has only $\sim10$ percent gas in the disc during the
first two encounters, due to gas consumption by star formation). As
the gas fraction in MBh is roughly twice the value of the MB run, the
cold gas component is able to absorb more of the kinetic impact energy
of the satellite. Therefore this process is also more efficient in the
MBh run than in the MB run. However, as this gas fraction is
comparable to the 40 per cent gas run of M10, it is not possible to
explain the final scale height by this effect alone, as M10 showed
that even an initial value of 40 percent cold gas leads to a final
scale height of $z_0 = 0.6\kpc$, and thus to considerable thickening
with respect to an isolated simulation. We note that the mass of the
cold gas and the SFR depend on the angular momentum of the hot gaseous
halo.

We thus find that the scale height of a MW-like galaxy can decrease
again after having been increased by a 1:10 merger, if cooling and
accretion from a gaseous halo are considered. With respect to
collisionless simulations ($\zo = 0.8\kpc$), the presence of 20
percent cold gas in the initial disc reduces the thickening by
$\sim25$ percent ($\zo = 0.7\kpc$), while the accretion of gas from
the hot halo further reduces the thickening. This final scale height
is similar to that of an isolated dissipationless simulation, such
that the 1:10 merger does not lead to a thicker disc in the
end. Because accretion of new gas is expected to be ubiquitous in a
cosmological context, we conclude that in order to retain thin discs,
like those observed in the universe, it is not necessary to have very
high initial gas fractions. Thus, a MW-like galaxy can experience a
1:10 merger without an overall increase of its scale height.

Although the new disc is very thin, the initial disc is thickened
by a considerable amount, possibly explaining the origin of thick
discs such as the one seen in our Galaxy. In the scenario presented
here, the thick disc is the old stellar disc that has been thickened in
a minor merger at $z\gta1$, while the thin disc is the new stellar disc
that reforms after this merger. In the MBh simulation the mean
stellar ages of the thin and thick discs are $\sim3.5\Gyr$ and $\gta6\Gyr$,
respectively. For the solar neighborhood the long-lived stars in the thin disc
are slightly older with a mean age of $\sim4\Gyr$. These results are
in good agreement with the observations by \citet{bensby2003} who 
analyze a sample of stars in the solar neighborhood and find
mean stellar ages of $4.9\pm2.0\Gyr$ and $11.2\pm4.3\Gyr$ for the thin
and thick disc, respectively. Using the same sample but a different method
of analyzing the data (i.e. different isochrones), \citet{feltzing2003} find
mean stellar ages of $6.1\pm2.0\Gyr$ and $12.1\pm3.8\Gyr$ for the two discs,
in agreement with our simulation. We note that the mean stellar age of the
new thin disc in our model, depends on the exact redshift of the minor merger.
If the satellite galaxy enters the main halo earlier than assumed in our simulation
($t_{\rm merge}>6\Gyr$) the mean stellar age of the new thin disc is larger, so the
thin disc can be older.

Scale heights for observed discs are usually quoted for an exponential profile,
rather than for a ${\rm sech}^2$ profile.
In order to compare the scale heights of the new thin disc and the thickened old disc
to observations, we compute the exponential scale heights of the final discs in our
MBh run. The exponential scale height of the new thin disc is $h_{\rm new}=0.35\kpc$
and that of the old thick disc is $h_{\rm old}=0.6\kpc$. These values are in very good
agreement with the observational constraints \citep[cf. the compilation presented in][]
{chang2011}.

To test the possibility for this scenario further, we compute the
velocity dispersions of the old thick disc and the new thin disc at the
solar radius. For the MBh run the velocity dispersion ellipsoid of the old thick disc is
$(\sigma_r,\sigma_\phi,\sigma_z)=(62,44,33)\kms$, while for new thin
disc it is $(\sigma_r,\sigma_\phi,\sigma_z)=(52,36,22)\kms$. These values
are in good agreement with observations by \citet{soubiran2003}, who find
$(\sigma_r,\sigma_\phi,\sigma_z)=(63,39,39)\kms$ for the thick disc and
$(\sigma_r,\sigma_\phi,\sigma_z)=(39,20,20)\kms$ for the thin disc.
Finally we note that the velocity dispersions in our simulation increase with stellar age,
in agreement with observations \citep{wielen1977,nordstrom2004,seabroke2007}.
For star with an age of $1\Gyr$ we get $(\sigma_r,\sigma_\phi,\sigma_z)=(41,29,15)\kms$,
while for stars with an age of $6\Gyr$ we get
$(\sigma_r,\sigma_\phi,\sigma_z)=(62,44,33)\kms$, consistent with the observed values.
This suggests that a scenario is very plausible, in which the thick disc results from the
thickening of an initial disc through a minor merger and the thin disc is formed after
this merger from accretion of gas from the halo and subsequent star formation.

\subsection{The effects of galactic winds}

\begin{figure}
\psfig{figure=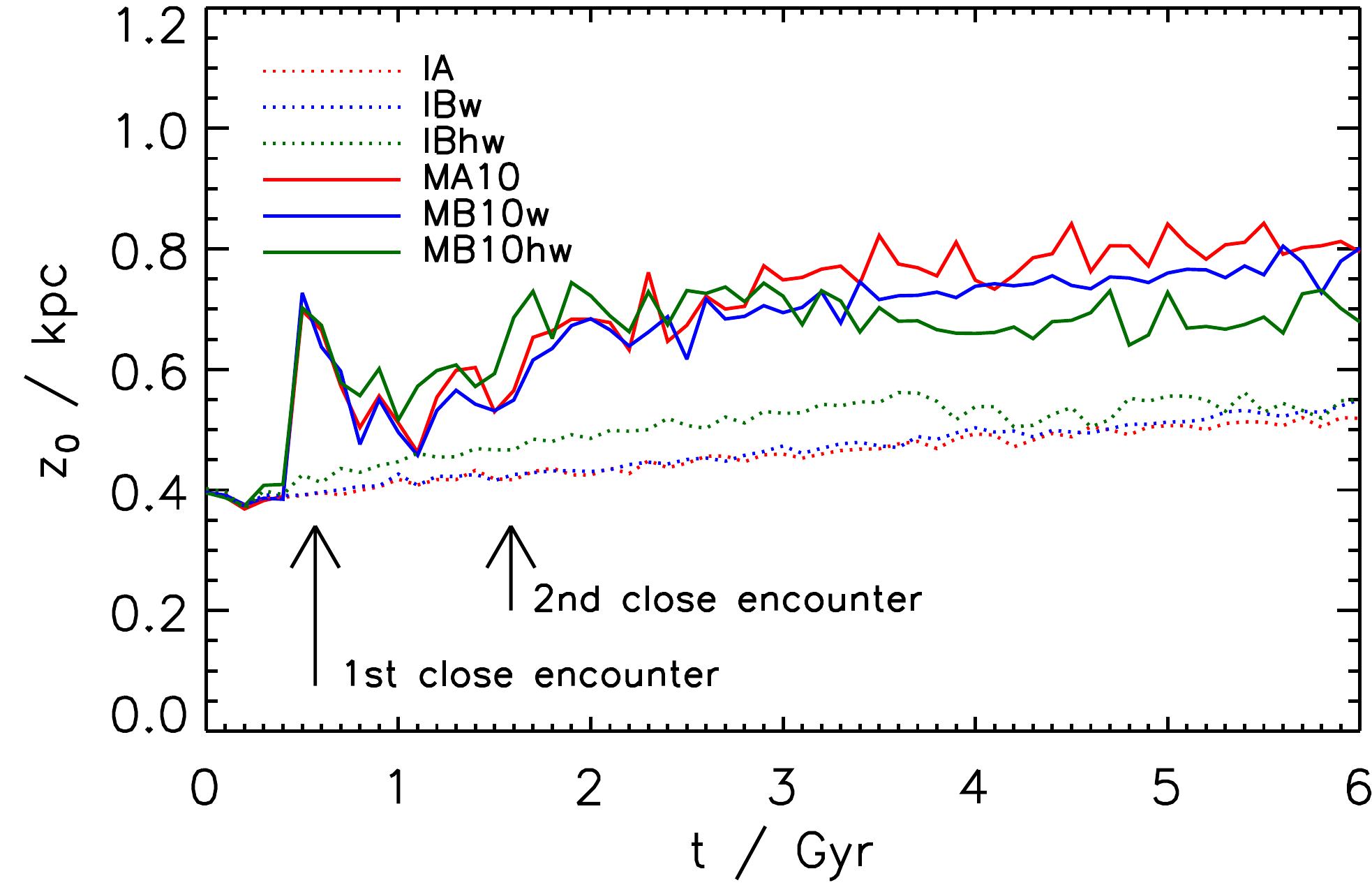,width=0.47\textwidth}
\caption{Evolution of the disc scale height for isolated and 1:10
  merger simulations with and without a gaseous halo for the case with
  galactic winds.}
\label{fig:z10whalo}
\end{figure}

In the following we study the evolution of the scale height during the
simulations that include galactic winds. The results are shown in
Figure \ref{fig:z10whalo} for isolated and 1:10 merger runs. When
winds are included, the isolated simulation with only 20 percent cold
gas in the disc (and no gaseous halo, IBw) shows more thickening than
in the case without winds (IB), leading to a final scale height that
is even slightly larger than in the dissipationless simulation
(IA). If the hot gaseous halo is included, the scale height increases even
more quickly than in the simulations without a hot halo, up to a value
of $\zo = 0.55\kpc$ at $t\sim4\Gyr$ after which it is roughly
constant. After $6\Gyr$ all isolated runs converge to the same scale
height.

The scale height of the 1:10 merger simulation with winds and without
a gaseous halo (MBw) shows considerably more thickening than the run
without winds (MB). With $\zo = 0.77\kpc$, it has almost the same
final scale height as the dissipationless run (MA). Even if the hot
gaseous halo is included in the simulation, the final scale height is
much larger than the final scale height of the isolated
dissipationless simulation with $\zo = 0.7\kpc$. In contrast to the
simulations without galactic winds, we find that a 1:10 merger leads to
a considerable increase of the scale height, producing discs with
final scale heights of $\zo\gta0.7\kpc$. We conclude that in order to
able to retain very thin discs, the efficiency of galactic winds has
to be rather low.

\begin{figure}
\psfig{figure=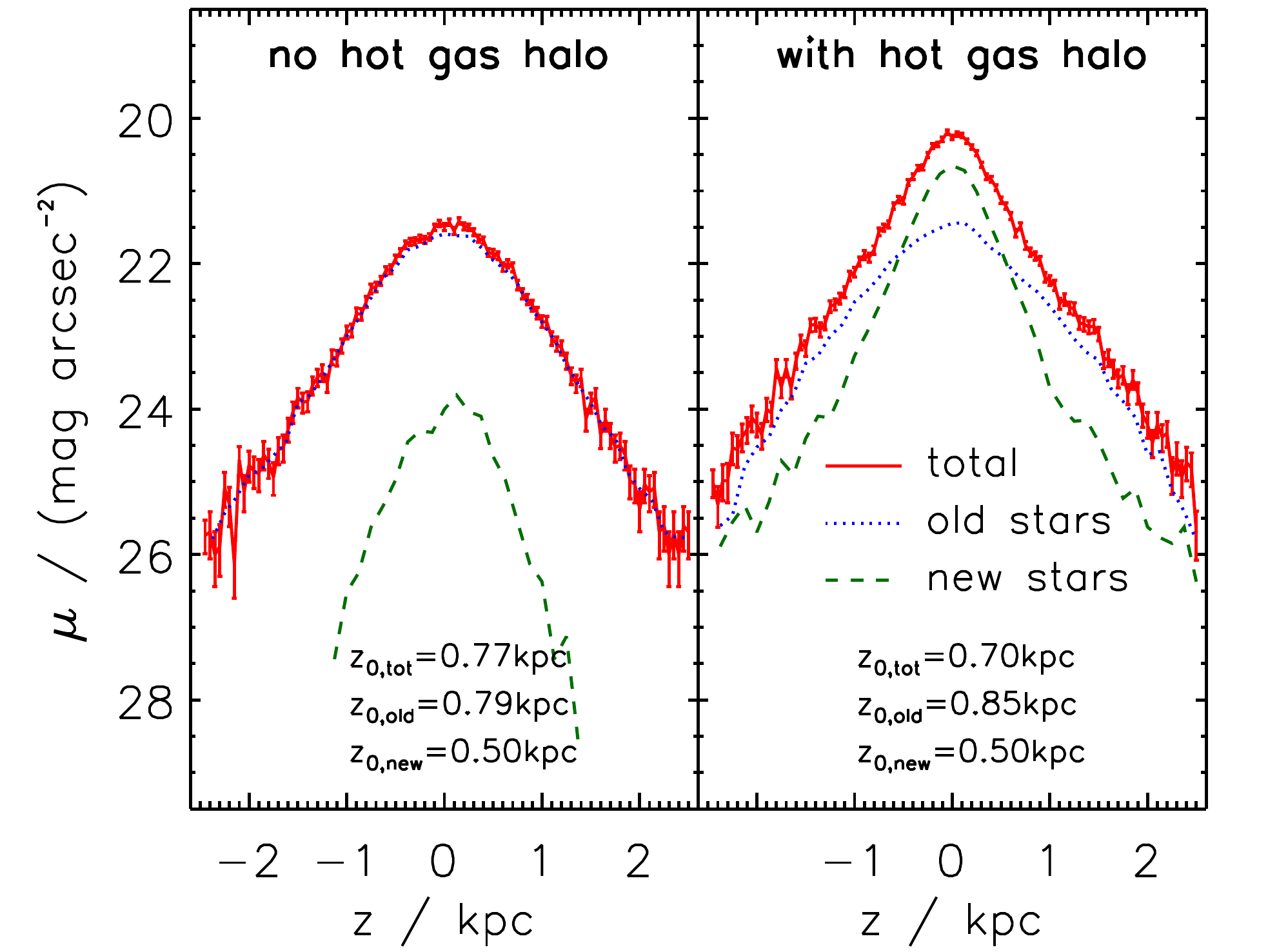,width=0.47\textwidth}
\caption{Final edge-on surface brightness profiles for the 1:10 merger
  simulations with winds at a projected radius of $8\kpc$ for old,
  new, and all stars.}
\label{fig:cprofw}
\end{figure}

We further investigate what is causing the increased scale heights
when galactic winds are considered. There are two possible mechanisms
that can lead to a thicker disc. One explanation is that if winds are
included the SFR is lowered, resulting in a less massive new disc at
the end of the simulation. Since the dominant effect in the run
without winds (but including a gaseous halo) is the reformation of a
massive new thin stellar disc that dominated the overall surface
density profile, a less massive disc will result in a smaller
contribution by the new disc to the total scale height. This leads to
a larger contribution due to the thickened old disc and thus to a
larger scale height. Another possibility that the winds actually alter
the mass distribution in the cold gas disc.  As the outflow of gas
exerts a pressure force on the cold gas component, the density in the
galactic plane is lowered, resulting in a shallower $\rho(z)$ profile
for the gaseous disc. This, in turn, leads to a shallower profile for
the new stellar disc and thus to a larger scale height for the total
stellar disc.

In order to study which of these effects is more important, we
determine the SFR and the stellar mass of the new stellar disc for the
simulations with winds and present the results in the right panels of
Figure \ref{fig:sfrminor}. Due to the winds, the SFR is indeed lowered
and has a constant value of $\sim5\Msunpyr$ for the simulations with a
hot gaseous halo. The final stellar mass of the new stellar disc is
therefore lower than the mass of the old stellar disc (even if hot gas
is included) and has only about half of the value that is obtained
without winds. As a result, the new stellar disc is less dominant in
the case with winds and the old stellar disc has a larger influence on
the total scale height.

This can also be seen in the edge-on projected surface brightness for
old, new and all stars. The profiles are presented in Figure
\ref{fig:cprofw} for the runs without hot gas (left panel) and
including the gaseous halo (right panel). Due to the very low SFR in
the case where the hot halo is neglected, the new stellar component
has much less mass than the old one. Therefore the surface brightness
is completely dominated by the old stellar disc and the total scale
height has a relatively large value of $z_0 = 0.77\kpc$. If the hot
gaseous halo is included, the new stellar disc is much more
massive. However, due to the winds, its final mass is still lower than
the mass of the old stellar disc. Therefore the new stellar component
dominates the surface brightness only close to the galactic plane and
the final total scale height ($z_0 = 0.7\kpc$) is much larger than for
the windless case. This shows that the effect caused by a lowered SFR
and stellar mass because of the winds has a large impact on the scale
height.

\begin{figure}
\psfig{figure=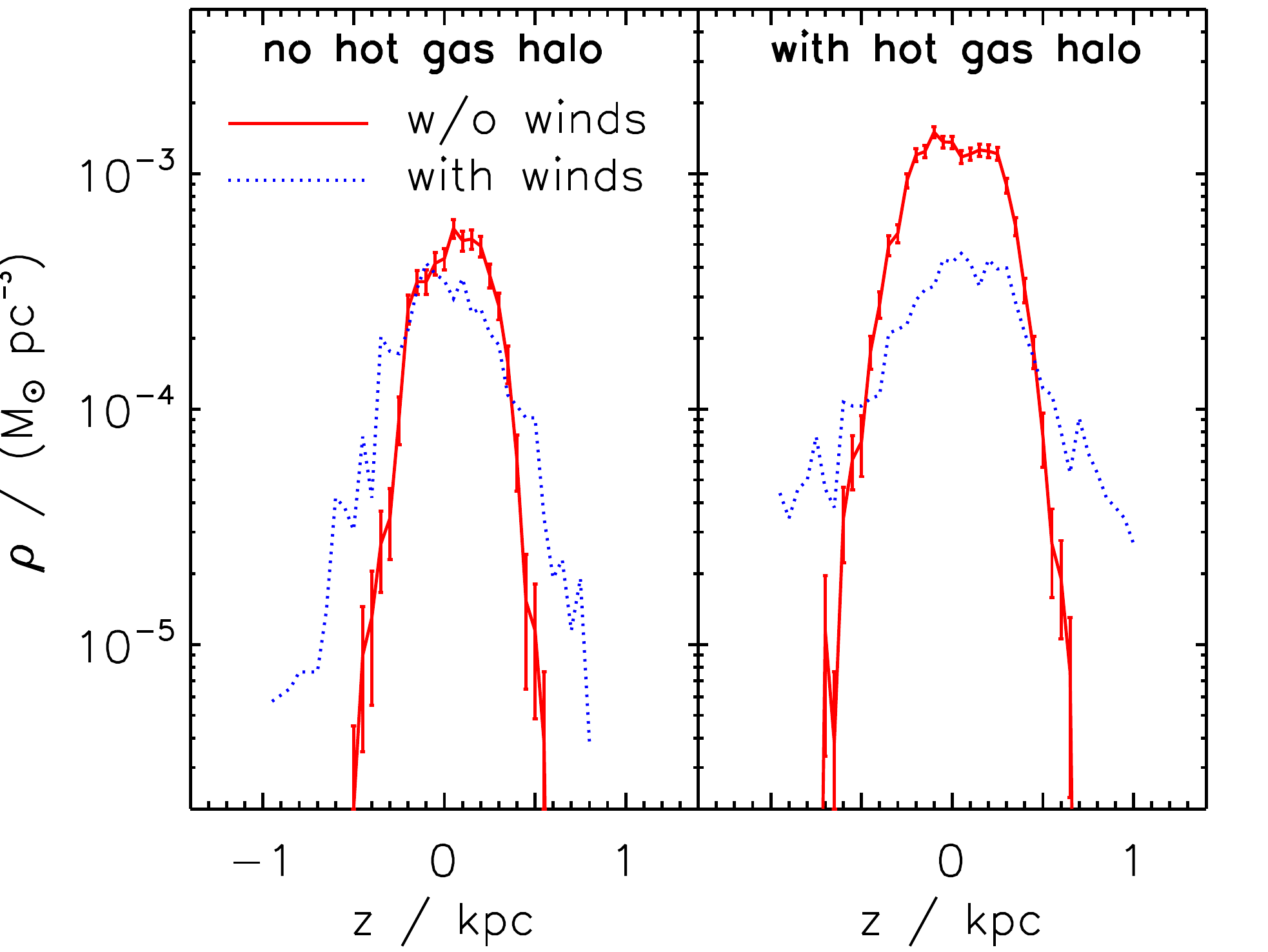,width=0.47\textwidth}
\caption{Final 3d density profiles for the cold gaseous discs in the
  1:10 merger simulations at a radius of $8\kpc$. The simulations without
  winds are given by the solid lines while the runs with winds are given by
  the dotted lines.}
\label{fig:cprofg}
\end{figure}

We also investigate whether the shallower gas potential due to the
pressure forces of the winds also affects the final scale height.  To
this end, we compute the 3d gas density as a function of distance from
the galactic plane at a radius of \Rsun. The resulting profiles are
shown in Figure \ref{fig:cprofg} for the runs without a hot component
(left panel) and for the runs with a gaseous halo (right panel). 
These results clearly show that in the simulations with galactic winds
the vertical density profile of the gas is much shallower than in the
runs without winds. Especially in the simulation with a gaseous halo,
the wind forces have pushed the cold gas to larger distances from the
plane, such that there is less material close to the disc plane. This, in
turn, also affects the profile of the new stellar discs that form from
these gaseous discs. Comparing the runs without winds (Figure
\ref{fig:cprofh}) to those with winds (Figure \ref{fig:cprofw}), we
find that the new stellar discs also have a shallower profile in the
runs with winds. As a result, the new disc in the run without hot gas
and with winds (MB10w) has a larger scale height of $z_{0,new} =
0.5\kpc$ compared to the run without winds (MB10) with $z_0 =
0.4\kpc$. Similarly, the run with a hot gaseous halo and with winds
(MB10hw) has a larger scale height of $z_{0,new} = 0.5\kpc$ compared
to the run without winds (MBh10) with $z_0 = 0.45\kpc$. Due to the
larger scale heights of the new stellar discs, the scale heights of
the total stellar discs are also larger.

The last question that has to be addressed for the simulations with
winds is, how much each of the two effects contribute to the larger
scale heights.  If we look at the evolution of the scale height in the
isolated run with winds and hot gas (IBhw), we see that \zo increases
more quickly than in the isolated simulation without winds.  Since there is no
merger event, the scale heights of the old stellar discs evolve
similarly for IBh, IBhw and even IA. This means that the increase of
\zo in IBhw can only be due to the evolution of the new stellar
disc. If the effect of a shallower gas profile due to the wind forces
was negligible compared to the effect of the lower SFR and stellar
mass, the scale height of the IBhw run should have a value between
those of the IBh run (maximum new stellar mass) and the IA run (no new
stellar mass). However, the scale height of the IBhw run is always
larger than the others. This indicates that the effect of a thicker
new stellar disc leads to this increase of \zo of the order of $\Delta
z_0\sim0.05\kpc$.  This result is in agreement with the finding that
the new stellar disc has a scale height that is roughly $\Delta
z_0\sim0.05\kpc$ larger when winds are considered. We thus conclude
that the scale height can increase by roughly $\Delta z_0\sim0.05\kpc$
due to the effect of the shallower potential caused by the wind
forces.

On the other hand, this value cannot explain the values of the scale
height in the 1:10 merger simulations. The difference in scale height
between the run with a hot halo and winds (MB10hw) and without winds
(MB10h) is of the order $\Delta z_0\sim0.2\kpc$. This means that this
large difference can only be explained by the lower stellar mass in
the run with winds, such that the thickened old disc still dominates
the overall surface brightness profile. We thus conclude that the more
important effect driving the increase of the scale height when winds
are considered comes from the lower contribution of the new thin
stellar disc due to the lowered SFR. The shallower profile of the
gaseous disc caused by the wind forces has a comparably minor
impact.

\subsection{Resolution study}
\label{sec:reso}

\begin{figure}
\psfig{figure=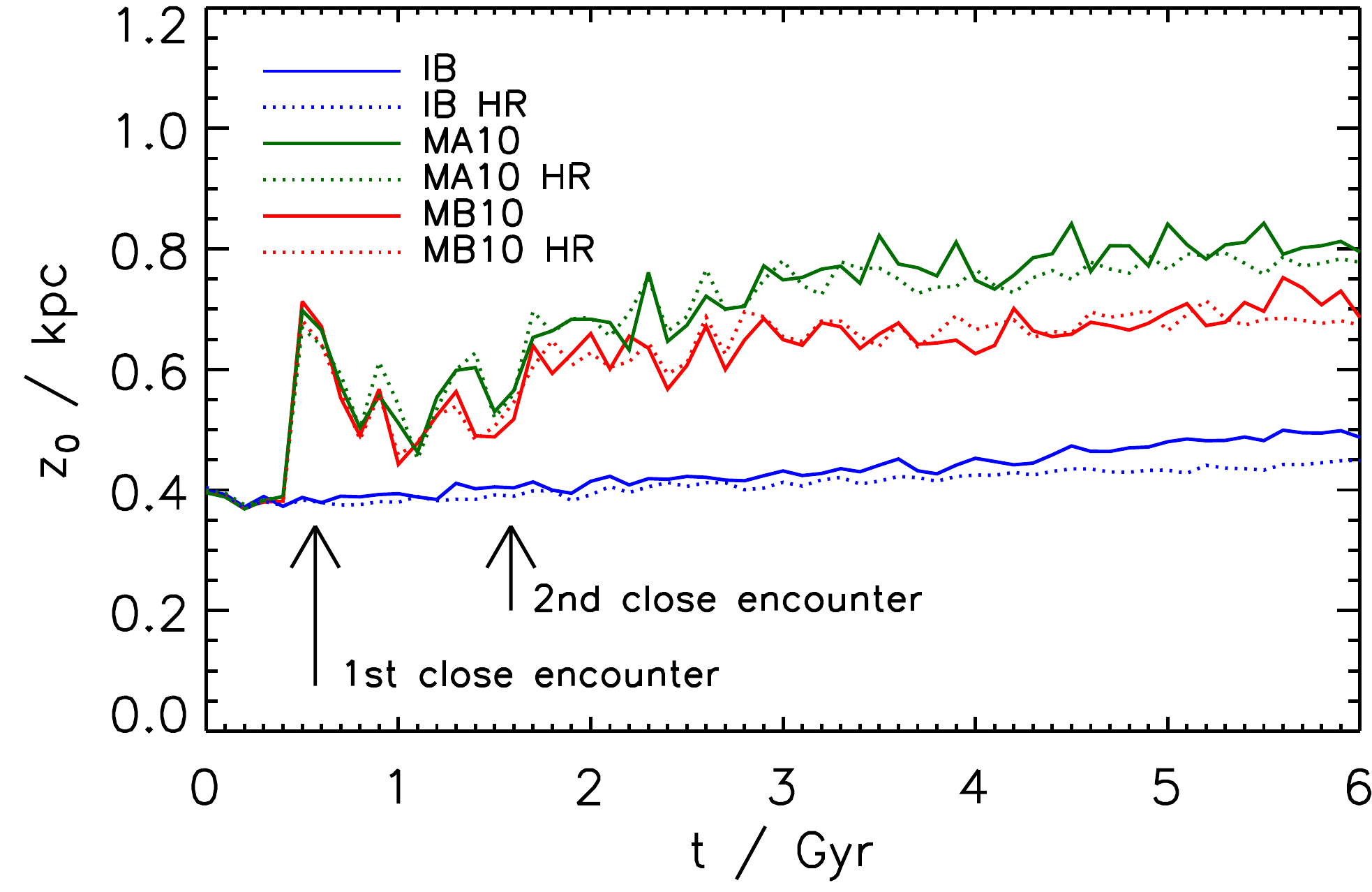,width=0.47\textwidth}
\caption{Comparison of the evolution of the scale height between the
  runs with standard resolution (solid lines) and resimulations with
  higher resolution (dotted lines) for the isolated and 1:10 merger
  simulations without a hot gaseous halo and without galactic winds.}
\label{fig:z10res}
\end{figure}

Since galactic discs are fragile systems, the morphology of numerical
realizations of discs can potentially be modified simply by numerical
effects, e.g. through bar formation or flaring. Therefore, it is
important to check how the evolution of the disc morphology changes if
the numerical resolution is altered. To this end we compare the
results obtained with our standard resolution to those obtained with
an increased resolution as employed by M10. These high resolution runs
have twice the number of dark matter and old stellar particles and
four times the number of gas particles in the initial conditions
compared to our standard resolution. This implies that the number of
new stellar particles is also four times higher.

First, the stability of the initial disc is a key point to be
addressed before attempting to study the effects of satellite
mergers. We compare the runs for the isolated disc with an initial
value of 20 percent cold gas in the disc, neglecting both the hot gas
and galactic winds (IB). The results are presented in Figure
\ref{fig:z10res}. While for the standard resolution run the scale
height increases to a final value of $z_0 = 0.5\kpc$ (from an initial
value of $z_0 = 0.4\kpc$), the high resolution run shows less
thickening with a final scale height of $z_0 = 0.45\kpc$. This shows
that for the isolated disc systems the stability is affected by
numerical resolution, although the difference between the standard and
the high resolution runs is small.  The total amount of thickening due
to numerical effects is very low compared to the thickening due to
satellite accretion events, even for the standard resolution. This
implies that any further increase in \zo is due to accretion events.

For the 1:10 merger simulations we do not find a difference between
the standard and high resolution runs. For both cases the scale height
is increased to the same value during the first two encounters and
evolves only slightly thereafter. Thus, if the discs are thickened
during a merger to scale heights of $z_0 \lta 0.6\kpc$, their further
evolution is the same for both resolutions. The reason for this is
that thicker discs are more robust to heating
\citep{kazantzidis2009}. This means that for our merger runs, our
chosen resolution is sufficient, i.e. we achieve numerical
convergence.

Finally, we compare the results for the standard and high resolution
runs for the 1:10 merger that includes both the hot gaseous halo and
galactic winds (MB10hw). The resulting evolution of the scale heights
is shown in Figure \ref{fig:z10res}. The results for the two runs
agree extremely well. For both simulations the scale height reaches a
maximum value after the second encounter and decreases again. We
therefore conclude that the effects of the hot gaseous halo and the
galactic winds are captured well with our standard resolution and that
also in this case, numerical convergence is achieved.

\begin{figure}
\psfig{figure=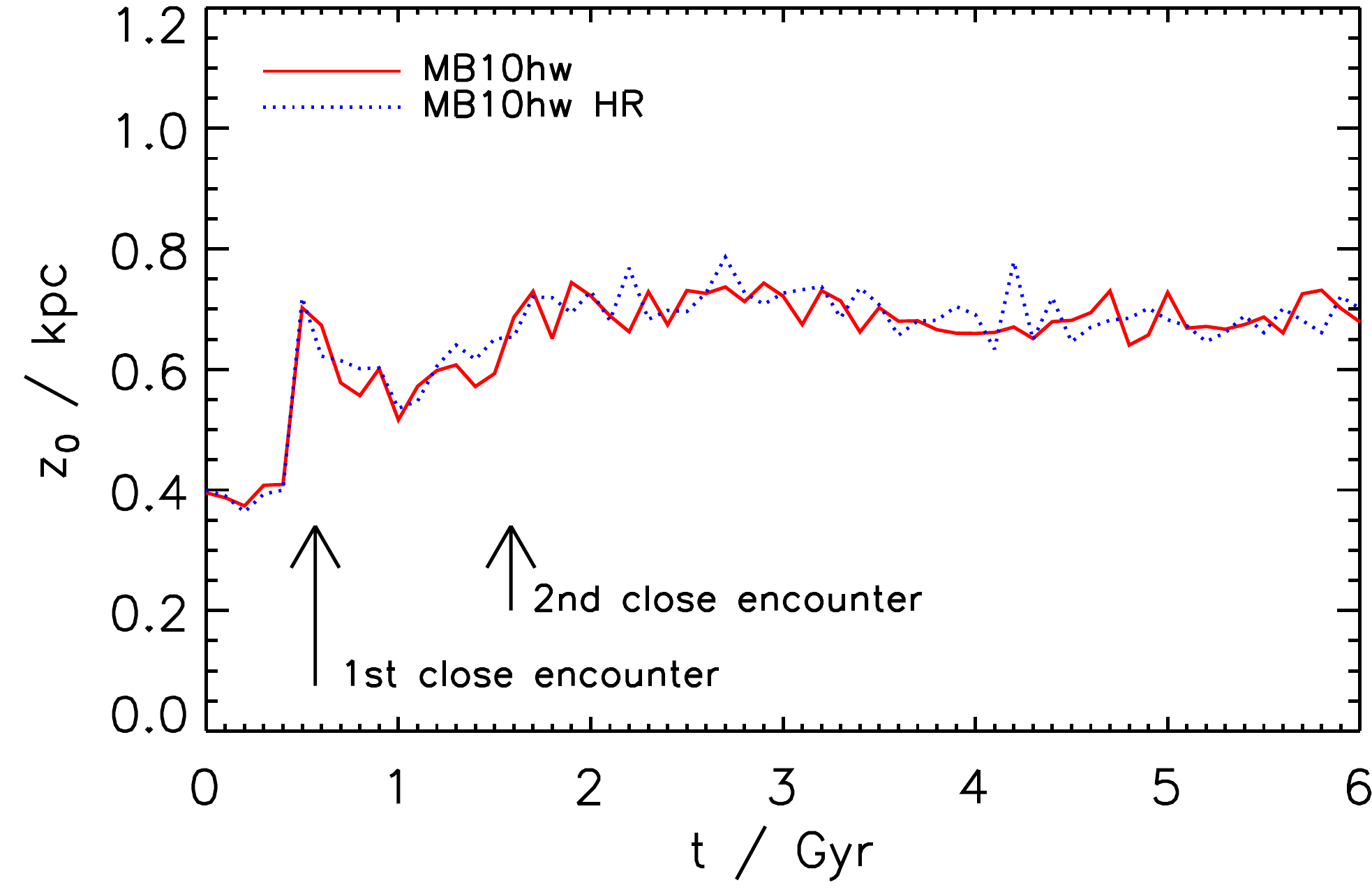,width=0.47\textwidth}
\caption{Comparison of the evolution of the scale height between the
  1:10 merger run with standard resolution (solid line) and the
  resimulation with high resolution (dotted line) for the simulations
  including a hot gaseous halo and galactic winds.}
\label{fig:z10hwres}
\end{figure}

\section{Results for the 1:5 merger}
\label{sec:res5}

In section \ref{sec:res10} we have shown that depending on the mass of
the hot gaseous halo and the efficiency of stellar driven winds or
other feedback processes, a MW-like galaxy can experience a 1:10
merger without being thickened with respect to an isolated
simulation. In this section we study the impact of a larger mass ratio
1:5 merger. As before we run merger simulations for a dissipationless
disc system (MA5), a disc galaxy that contains an initial value of 20
percent cold gas in the disc (MB5) and a system that also includes a
hot gaseous halo (MB5h). 

\begin{figure}
\psfig{figure=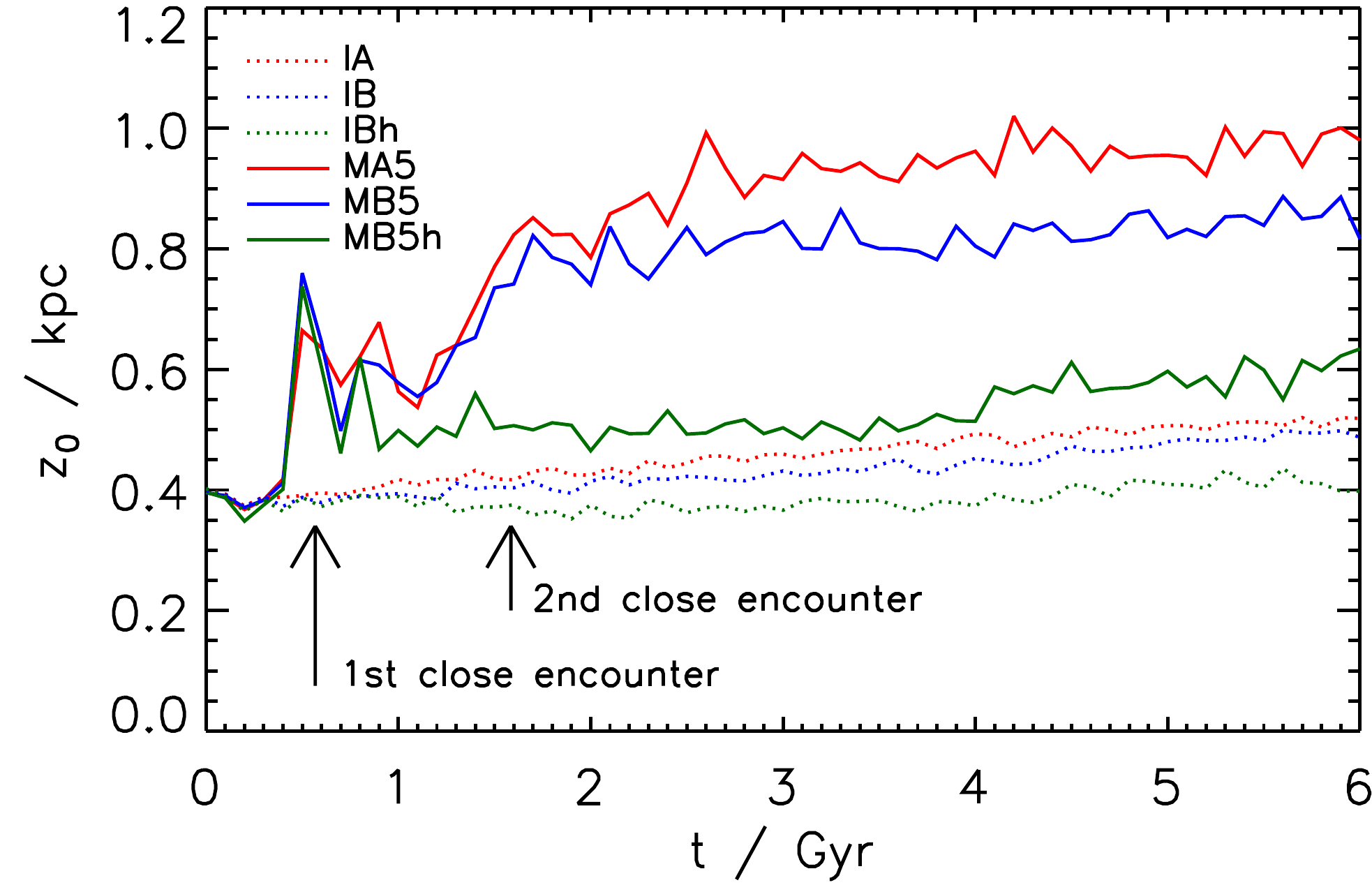,width=0.47\textwidth}
\caption{Evolution of the disc scale height for isolated and 1:5
  merger simulations with and without a gaseous halo for the case
  without galactic winds.}
\label{fig:z5halo}
\end{figure}

The resulting evolution of the scale height for the 1:5 merger
simulations without galactic winds is shown in Figure
\ref{fig:z5halo}. In the run without gas, the disc is strongly
thickened with a final scale height of $z_0 = 1\kpc$. This implies
that the thin stellar disc is completely destroyed and only a thick
disc is left, leading to the conclusion that the thin disc of a
MW-like galaxy cannot survive a dissipationless 1:5 merger. If the
initial disc contains 20 percent cold gas, the thickening is
considerably reduced with a final scale height of $z_0 = 0.85\kpc$.
Similar to the 1:10 merger, the cold gas in the disc absorbs some of
the impact energy of the satellite and is able to radiate it away,
such that the amount of thickening during the first two encounters is
reduced. After the second encounter, the scale height is relatively
constant, so the effects of the new stellar disc forming from the cold
gas are very small. If the hot gas in the halo is also considered, the
amount of thickening is greatly reduced. The final scale height of
$z_0 = 0.6\kpc$ is only slightly larger than that in the isolated
simulation. The resulting system thus has a scale height that is
consistent with that of observed MW-like galaxies. We therefore
conclude that if enough hot gas can cool and form stars, a
MW-like system is able to have a thin disc even after a 1:5 merger.

If we also consider the galactic winds, the final scale heights of the
simulations with gas increase with respect to the windless case, in
agreement with the results we obtained for the 1:10 merger
simulations. The run that includes only gas in the disc (MB5w) shows
the same evolution of the scale height as the dissipationless run. The
reason for this is the very small SFR and resulting mass of the new
stellar disc, such that its contribution to the final surface
brightness profile is negligible. Moreover, the amount of cold gas is
reduced considerably by the winds, so that the lower mass gas disc
cannot absorb impact energy. The simulation with winds and only gas in
the disc is therefore similar to a dissipationless run.

If the hot gaseous halo is included (MB5hw), the scale height
increases during the first two encounter and then decreases again to a
final value of $z_0 = 0.8\kpc$. Again, due to the winds, the gas
fraction during the first two encounters is reduced, such that the
amount of impact energy that can be absorbed is very small. As a
result, this simulation behaves in a similar manner as a collisionless
run at the beginning of the run. However, after the disc has been
thickened, the mass of the new stellar disc grows and starts to affect
the total surface brightness profile. As the new disc becomes
comparable in mass to the old stellar disc, the total scale height
decreases again. However, as the stellar mass of the new disc is much
lower than in the windless case, the final scale height is still much
higher. Nevertheless, even if galactic winds are operating, a MW-like
galaxy can experience a 1:5 merger that leads to a system that is in
the observable range of scale heights. We conclude that, if there is
enough hot gas available in the halo and the wind efficiencies are
very high, a disc system of the size of the MW can survive a 1:5
merger, without being in conflict with observational constraints.

\section{Conclusions and discussion}
\label{sec:conc}

We investigated the role of a cooling gaseous halo in minor merger
simulations using the parallel TreeSPH-code {\sc GADGET-2}. For this
we followed M11 and extended the initial conditions to include a hot
gaseous halo (in addition to a dark matter halo, a stellar bulge and a
disc consisting of stars and cold gas). We adopted the observationally
motivated $\beta$-profile and demanded that the halo be in hydrostatic
equilibrium. Furthermore the gaseous halo is rotating around the spin
axis of the disc. We have fixed its angular momentum by requiring that
the specific angular momentum of the gas halo is a multiple $\alpha$
of the specific angular momentum of the dark matter halo and treated
$\alpha$ as a free parameter. We then fixed the value of $\alpha$ by
requiring our simulations to reproduce the observed size-mass relation
for galactic discs (see M11). We fixed the mass of the gaseous halo by
requiring that the baryonic fraction within the virial radius is 85
percent of the universal one. As discussed in detail in M11, our
addopted gaseous haloes are not in conflict with X-ray observations of
MW-like spiral galaxies.

\begin{figure}
\psfig{figure=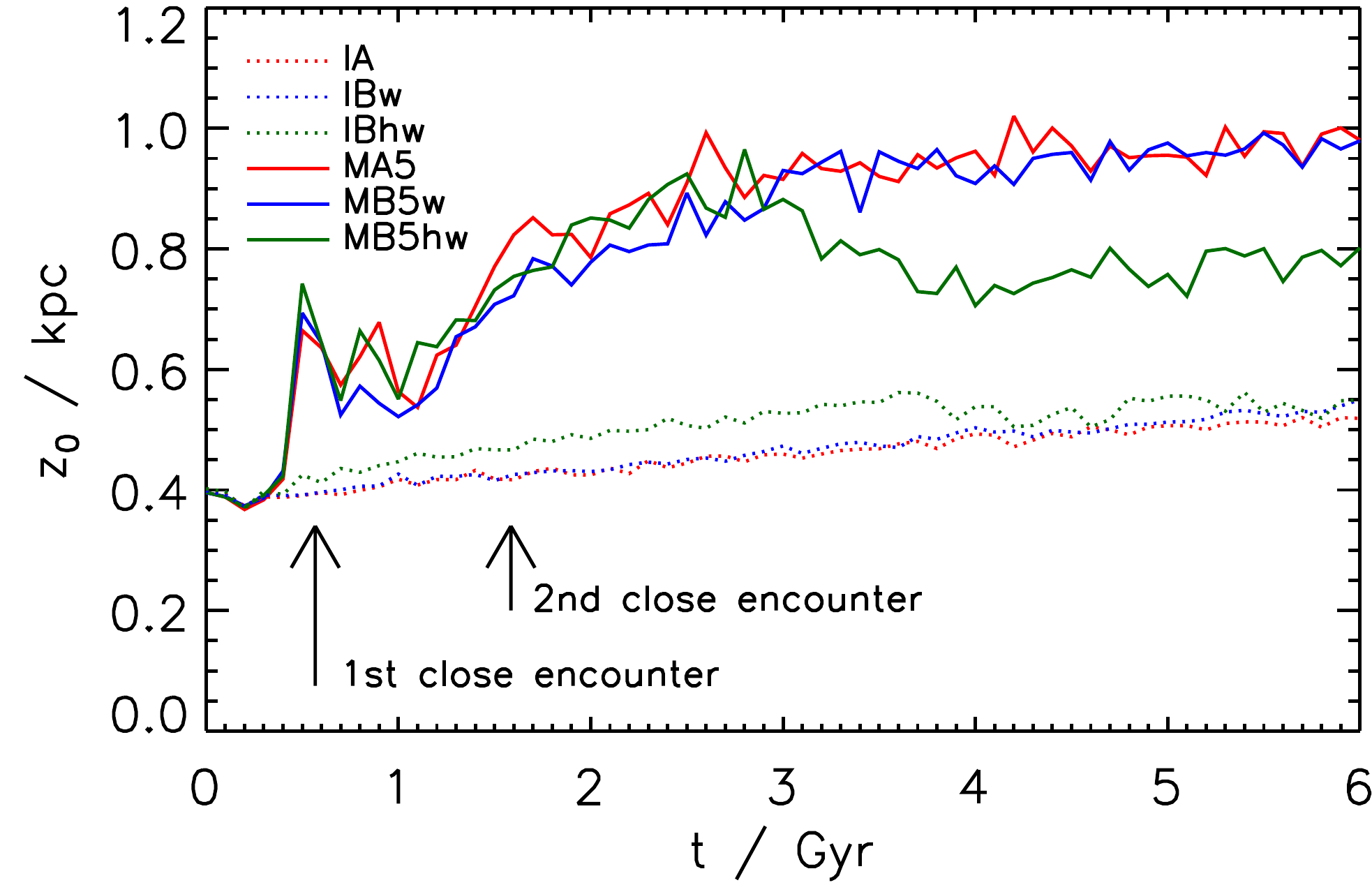,width=0.47\textwidth}
\caption{Evolution of the disc scale height for isolated and 1:5
  merger simulations with and without a gaseous halo for the case with
  galactic winds.}
\label{fig:z5whalo}
\end{figure}

We have revisited the question of whether thin discs like the one
observed in the MW can survive minor merger events. In collisionless
simulations of a 1:10 merger, such a disc is thickened considerably
($\zo=0.8\kpc$), while M10 have shown that if the cold gas component
in the disc is considered, the thickening due to the merger can be
reduced ($\zo=0.7\kpc$). The reason for that is the absorption of
impact energy by the gas.  Here, we have extended this study in order
to also consider the hot gas component in the halo and found that the
thickening with respect to collisionless simulations is further
reduced, leading to a similar scale height as in an isolated run
($\zo=0.5\kpc$). This is due to the cooling and accretion of new cold
gas which leads to the formation of a new thin disc. Similar to the
conclusions reached in major merger simulations, we find that high
disc cold gas fractions of $\sim40$ percent are not needed in order to
form realistic galaxies with properties as found in observations. The
subsequent cooling of the gaseous halo compensates for lower initial
cold gas fractions. We conclude that if a considerable amount of hot
gas is available in the halo, a MW-like disc system can survive a 1:10
merger as a new thin disc reforms after the merger.

The regrowing thin disc can reduce the final scale height in two
ways. First, the additional potential can cause heated stars to
recontract towards the disc plane. Second, the new thin stellar disc
is more massive than the old disc and therefore dominates the surface
brightness profile leading to a lower overall scale height. We have
shown that in our simulations the old stellar disc does not
recontract, i.e. its scale height is not lower than in a merger
simulation without gas. Therefore this effect cannot explain the small
scale height of the total disc.  Instead, we have demonstrated that
the massive new stellar disc dominates the surface brightness profile
resulting in a smaller overall scale height. Due to the subsequent
cooling from the halo and star formation, the new disc is much more
massive than in simulations that only have gas in the disc.

In the scenario presented in this work, the thick disc is the old stellar disc that has been
thickened in a minor merger at $z\gta1$, while the thin disc is the new stellar disc
that reforms after this merger. In the simulation with a hot gaseous halo, the mean stellar
ages of the thin and thick discs in the solar neighborhood are $\sim4\Gyr$ and $\gta6\Gyr$,
respectively, in agreement with measurements in the MW. We found an exponential scale
height for the new thin disc of $h_{\rm new}=0.35\kpc$ and for the old thick disc of
$h_{\rm old}=0.6\kpc$ in very good agreement with the values obtained for the MW.
Moreover we measured a velocity dispersion ellipsoid for the old thick disc of
$(\sigma_r,\sigma_\phi,\sigma_z)=(62,44,33)\kms$ and for new thin disc we found
$(\sigma_r,\sigma_\phi,\sigma_z)=(52,36,22)\kms$. The good agreement with observed
values suggests that this scenario for the formation of the thin and thick discs is very
plausible.

We have addressed the effects of galactic winds on the results of our
simulations and found that if winds are included, the final scale
heights are considerably larger. There are two effects the winds can
have on the evolution of the scale height. On the one hand, the
efficiency of gas cooling from the hot component and star formation
partially depend on the feedback mechanism implemented in the
simulations. If the winds are very efficient, the cooling rate and the
SFR are lowered, resulting in a less massive new thin disc. In this
case the thickened old stellar disc still dominates the surface
brightness and the total scale height is larger. On the other hand,
winds can exert a pressure force on the cold gas in the disc. This
leads to an expansion of the cold gas and thus to a shallower
profile. This in turn, results in a thicker new stellar disc and thus
in a larger overall scale height.

To test how much each of these processes contributes to the increased
scale heights when winds are present, we have performed a series of
simulations with and without winds. In order to check whether the
pressure forces of the winds have an effect on the profile of the gas,
we computed the 3d density profile of the cold gas for the case with
and without winds. We found that the cold gas profile $\rho(z)$ is
indeed much shallower when winds are included. This, in turn, affects
the thickness of the new stellar disc that forms from this gaseous
disc. The scale height of the new disc is therefore roughly
$\sim0.05-0.1\kpc$ larger when winds are present. Moreover the
galactic winds reduce the SFR by a factor of $\sim2$, resulting in a
new stellar disc that is less massive than the old one. The new disc
then only dominates the surface brightness very close to the disc
plane, and the scale height is fixed by the thicker old disc. This
effect can lead to a scale height that is roughly $\sim0.2\kpc$ larger
when winds are operating. We therefore conclude that lowered SFR has a
larger impact on the evolution of the scale height than the pressure
force exerted by the winds.

For all merger simulations with galactic winds we used the `constant
wind' model of \citet{springel2003}, where the mass-loss rate driven
by the winds is proportional to the SFR and the wind speed is
constant. 
However, this simple assumption does not lead to very good
agreement with observations such as the stellar mass vs. metallicity
relationship for galaxies. Wind models in which the mass loading
factor and wind velocity scale with galaxy internal velocity,
motivated by the expected scalings for momentum driven winds, produce
much better agreement with these observational scaling relations
\citep[]{oppenheimer2006}.  However, the
wind speeds and the mass loading factor for MW-like galaxies in this
model do not vary much after $z=1$. Since the minor mergers that
increase the disc scale height are expected to occur at low redshift
and there is no star forming component associated with the satellite,
we are fairly confident that a more sophisticated wind model would
leave our qualitative results unchanged.

We have further tested the numerical convergence of our models. To do
this, we have employed simulations with twice the resolution in dark
matter and old stars and four times the resolution in gas and new
stars. Comparing our standard resolution simulations to the high
resolution ones we found no difference in the evolution of the scale
height during 1:10 mergers and only a small difference in the isolated
runs. This shows that numerical convergence has been achieved and that
any further increase in \zo with respect to an isolated simulation is
caused by a merger event.

The simulations for the 1:10 merger showed that a MW-like galaxy can
have a thin disc after such an event without showing excessive heating if enough hot
gas is available from which new gaseous and stellar discs can
reform. In the light of this, we addressed the question whether such a
galaxy can also have a thin disc after a 1:5 merger. To this end, we performed a
suite of simulations with and without a hot gas component and galactic
winds. We found that in a collisionless 1:5 merger simulation, the
disc is thickened to a final value of $\zo=1\kpc$, i.e. the thin disc
is destroyed. If an initial value of 20 percent cold gas is included
in the disc, this thickening can be slightly reduced to a final value
of $\zo=0.85\kpc$. However, if the hot gaseous halo is included, the
final scale height is much smaller ($\zo=0.6\kpc$) and only slightly
larger than in an isolated run. The reasons for this reduced
thickening are the same as for the 1:10 merger. If both hot gas and
galactic winds are considered, the final scale height after a 1:5
merger is $\zo=0.8\kpc$. Our simulations show that a MW-like galaxy
can have a thin and a thick disc after such a merger if there is enough hot gas
available in the halo and if the wind efficiency is low.

We found that the final scale heights in our simulations depend on the
mass of the hot gaseous halo, the efficiency of the winds and the
merger ratio. The values of the scale height in our post-merger sample
are in the range $0.5\lta\zo\lta1.0\kpc$. In the light of these
results, we can assess whether our simulated systems are in agreement
with observational findings, or whether the scale heights in our
simulations present a problem for the CDM theory. Obtaining
statistically unbiased observational measurements of scale heights for
a complete sample of galaxies is difficult, due to small sample sizes,
dust extinction and inclination effects. However, there are a
considerable number of studies focusing on the vertical structure of
galaxies
\citep{shaw1989,shaw1990,grijs1996,grijs1997,pohlen2000,kregel2002,bizyaev2004}. These
studies suggest that the majority of observed MW-like galaxies have
scale heights in the range of $0.6\lta\zo\lta1.0\kpc$. The statistical
studies by \citet{schwarzkopf2000} and \citet{yoachim2006} find that
the distribution of scale heights for disc galaxies of the same mass
as the MW is between 0.4 and 1.2 kpc with a maximum around
$\sim0.6-0.8\kpc$. The scale heights of our sample fall into this
range. We therefore conclude that our simulations are in good
agreement with observational findings.

This work underlines once again the importance of the inclusion of a
dissipational component in galaxy merger studies. This component, even
though subdominant from a gravitational point of view, is
able to strongly modify the dynamical response of the stellar disc
during minor mergers, making them more resistant against heating.
Once both cold and hot gas are taken into account, claimed tensions
between the hierarchical model and the abundance of thin disks in the
local Universe appear to be relaxed.

\section*{Acknowledgements} 

%
We thank Chao Liu for helpful comments on the observed properties of
the thin and thick discs in the MW and Shy Genel for helpful comments
on merger statistics. Furthermore we thank Simon White
for enlightening discussions and useful comments on this work.
The numerical simulations used in this work were performed
on the PIA and THEO clusters of the Max-Planck-Institut f\"ur Astronomie and
on the OPA cluster of the Max-Planck-Institut f\"ur Astrophysik.
AVM acknowledges funding by Sonderforschungsbereich SFB 881
``The Milky Way System'' (subproject A1) of the German Research Foundation (DFG).


\bibliographystyle{mn2e}
\bibliography{moster2011}

\begin{thebibliography}{}

\bibitem[\protect\citeauthoryear{{Abadi}, {Navarro}, {Steinmetz} \&
  {Eke}}{{Abadi} et~al.}{2003}]{abadi2003}
{Abadi} M.~G.,  {Navarro} J.~F.,  {Steinmetz} M.,    {Eke} V.~R.,  2003, \apj,
  597, 21

\bibitem[\protect\citeauthoryear{{Bensby}, {Feltzing} \&
  {Lundstr{\"o}m}}{{Bensby} et~al.}{2003}]{bensby2003}
{Bensby} T.,  {Feltzing} S.,    {Lundstr{\"o}m} I.,  2003, \aap, 410, 527

\bibitem[\protect\citeauthoryear{{Benson}, {Lacey}, {Frenk}, {Baugh} \&
  {Cole}}{{Benson} et~al.}{2004}]{benson2004}
{Benson} A.~J.,  {Lacey} C.~G.,  {Frenk} C.~S.,  {Baugh} C.~M.,    {Cole} S.,
  2004, \mnras, 351, 1215

\bibitem[\protect\citeauthoryear{{Bizyaev} \& {Kajsin}}{{Bizyaev} \&
  {Kajsin}}{2004}]{bizyaev2004}
{Bizyaev} D.,  {Kajsin} S.,  2004, \apj, 613, 886

\bibitem[\protect\citeauthoryear{{Bournaud}, {Jog} \& {Combes}}{{Bournaud}
  et~al.}{2005}]{bournaud2005}
{Bournaud} F.,  {Jog} C.~J.,    {Combes} F.,  2005, \aap, 437, 69

\bibitem[\protect\citeauthoryear{{Bower}, {Benson}, {Malbon}, {Helly}, {Frenk},
  {Baugh}, {Cole} \& {Lacey}}{{Bower} et~al.}{2006}]{bower2006}
{Bower} R.~G.,  {Benson} A.~J.,  {Malbon} R.,  {Helly} J.~C.,  {Frenk} C.~S.,
  {Baugh} C.~M.,  {Cole} S.,    {Lacey} C.~G.,  2006, \mnras, 370, 645

\bibitem[\protect\citeauthoryear{{Brook}, {Kawata}, {Gibson} \&
  {Freeman}}{{Brook} et~al.}{2004}]{brook2004}
{Brook} C.~B.,  {Kawata} D.,  {Gibson} B.~K.,    {Freeman} K.~C.,  2004, \apj,
  612, 894

\bibitem[\protect\citeauthoryear{{Cavaliere} \& {Fusco-Femiano}}{{Cavaliere} \&
  {Fusco-Femiano}}{1976}]{cavaliere1976}
{Cavaliere} A.,  {Fusco-Femiano} R.,  1976, \aap, 49, 137

\bibitem[\protect\citeauthoryear{{Chang}, {Ko} \& {Peng}}{{Chang}
  et~al.}{2011}]{chang2011}
{Chang} C.-K.,  {Ko} C.-M.,    {Peng} T.-H.,  2011, arXiv:1107.3884

\bibitem[\protect\citeauthoryear{{Cox}, {Dutta}, {Di Matteo}, {Hernquist},
  {Hopkins}, {Robertson} \& {Springel}}{{Cox} et~al.}{2006}]{cox2006}
{Cox} T.~J.,  {Dutta} S.~N.,  {Di Matteo} T.,  {Hernquist} L.,  {Hopkins}
  P.~F.,  {Robertson} B.,    {Springel} V.,  2006, \apj, 650, 791

\bibitem[\protect\citeauthoryear{{de Grijs}, {Peletier} \& {van der Kruit}}{{de
  Grijs} et~al.}{1997}]{grijs1997}
{de Grijs} R.,  {Peletier} R.~F.,    {van der Kruit} P.~C.,  1997, \aap, 327,
  966

\bibitem[\protect\citeauthoryear{{de Grijs} \& {van der Kruit}}{{de Grijs} \&
  {van der Kruit}}{1996}]{grijs1996}
{de Grijs} R.,  {van der Kruit} P.~C.,  1996, \aaps, 117, 19

\bibitem[\protect\citeauthoryear{{Eke}, {Navarro} \& {Frenk}}{{Eke}
  et~al.}{1998}]{eke1998}
{Eke} V.~R.,  {Navarro} J.~F.,    {Frenk} C.~S.,  1998, \apj, 503, 569

\bibitem[\protect\citeauthoryear{{Fakhouri}, {Ma} \&
  {Boylan-Kolchin}}{{Fakhouri} et~al.}{2010}]{fakhouri2010}
{Fakhouri} O.,  {Ma} C.-P.,    {Boylan-Kolchin} M.,  2010, \mnras, 406, 2267

\bibitem[\protect\citeauthoryear{{Feltzing}, {Bensby} \&
  {Lundstr{\"o}m}}{{Feltzing} et~al.}{2003}]{feltzing2003}
{Feltzing} S.,  {Bensby} T.,    {Lundstr{\"o}m} I.,  2003, \aap, 397, L1

\bibitem[\protect\citeauthoryear{{Font}, {Navarro}, {Stadel} \& {Quinn}}{{Font}
  et~al.}{2001}]{font2001}
{Font} A.~S.,  {Navarro} J.~F.,  {Stadel} J.,    {Quinn} T.,  2001, \apj, 563,
  L1

\bibitem[\protect\citeauthoryear{{Gauthier}, {Dubinski} \& {Widrow}}{{Gauthier}
  et~al.}{2006}]{gauthier2006}
{Gauthier} J.-R.,  {Dubinski} J.,    {Widrow} L.~M.,  2006, \apj, 653, 1180

\bibitem[\protect\citeauthoryear{{Genel}, {Bouch{\'e}}, {Naab}, {Sternberg} \&
  {Genzel}}{{Genel} et~al.}{2010}]{genel2010}
{Genel} S.,  {Bouch{\'e}} N.,  {Naab} T.,  {Sternberg} A.,    {Genzel} R.,
  2010, \apj, 719, 229

\bibitem[\protect\citeauthoryear{{Gingold} \& {Monaghan}}{{Gingold} \&
  {Monaghan}}{1977}]{gingold1977}
{Gingold} R.~A.,  {Monaghan} J.~J.,  1977, \mnras, 181, 375

\bibitem[\protect\citeauthoryear{{Haardt} \& {Madau}}{{Haardt} \&
  {Madau}}{1996}]{Haardt1996}
{Haardt} F.,  {Madau} P.,  1996, \apj, 461, 20

\bibitem[\protect\citeauthoryear{{Hansen}, {Macci{\'o}}, {Romano-Diaz},
  {Hoffman}, {Br{\"u}ggen}, {Scannapieco} \& {Stinson}}{{Hansen}
  et~al.}{2010}]{hansen2010}
{Hansen} S.~H.,  {Macci{\'o}} A.~V.,  {Romano-Diaz} E.,  {Hoffman} Y.,
  {Br{\"u}ggen} M.,  {Scannapieco} E.,    {Stinson} G.~S., 2011, \apj, 734, 62

\bibitem[\protect\citeauthoryear{{Hernquist}}{{Hernquist}}{1990}]{hernquist1990}
{Hernquist} L.,  1990, \apj, 356, 359

\bibitem[\protect\citeauthoryear{{Hernquist}}{{Hernquist}}{1992}]{hernquist1992}
{Hernquist} L.,  1992, \apj, 400, 460

\bibitem[\protect\citeauthoryear{{Huang} \& {Carlberg}}{{Huang} \&
  {Carlberg}}{1997}]{sellwood1997}
{Huang} S.,  {Carlberg} R.~G.,  1997, \apj, 480, 503

\bibitem[\protect\citeauthoryear{{Johansson}, {Naab} \& {Ostriker}}{{Johansson}
  et~al.}{2009}]{johansson2009b}
{Johansson} P.~H.,  {Naab} T.,    {Ostriker} J.~P.,  2009, \apjl, 697, L38

\bibitem[\protect\citeauthoryear{{Jones} \& {Forman}}{{Jones} \&
  {Forman}}{1984}]{jones1984}
{Jones} C.,  {Forman} W.,  1984, \apj, 276, 38

\bibitem[\protect\citeauthoryear{{Katz}, {Weinberg} \& {Hernquist}}{{Katz}
  et~al.}{1996}]{katz1996}
{Katz} N.,  {Weinberg} D.~H.,    {Hernquist} L.,  1996, \apjs, 105, 19

\bibitem[\protect\citeauthoryear{{Kauffmann}, {White} \&
  {Guiderdoni}}{{Kauffmann} et~al.}{1993}]{kauffmann1993}
{Kauffmann} G.,  {White} S.~D.~M.,    {Guiderdoni} B.,  1993, \mnras, 264, 201

\bibitem[\protect\citeauthoryear{{Kazantzidis}, {Bullock}, {Zentner},
  {Kravtsov} \& {Moustakas}}{{Kazantzidis} et~al.}{2008}]{kazantzidis2008}
{Kazantzidis} S.,  {Bullock} J.~S.,  {Zentner} A.~R.,  {Kravtsov} A.~V.,
  {Moustakas} L.~A.,  2008, \apj, 688, 254

\bibitem[\protect\citeauthoryear{{Kazantzidis}, {Zentner}, {Kravtsov},
  {Bullock} \& {Debattista}}{{Kazantzidis} et~al.}{2009}]{kazantzidis2009}
{Kazantzidis} S.,  {Zentner} A.~R.,  {Kravtsov} A.~V.,  {Bullock} J.~S.,
  {Debattista} V.~P., 2009, \apj, 700, 1896 

\bibitem[\protect\citeauthoryear{{Kennicutt}
  Jr.}{{Kennicutt}}{1998}]{kennicutt1998}
{Kennicutt} Jr. R.~C.,  1998, \apj, 498, 541

\bibitem[\protect\citeauthoryear{{Klypin}, {Gottl{\"o}ber}, {Kravtsov} \&
  {Khokhlov}}{{Klypin} et~al.}{1999}]{klypin1999}
{Klypin} A.,  {Gottl{\"o}ber} S.,  {Kravtsov} A.~V.,    {Khokhlov} A.~M.,
  1999, \apj, 516, 530

\bibitem[\protect\citeauthoryear{{Kregel}, {van der Kruit} \& {de
  Grijs}}{{Kregel} et~al.}{2002}]{kregel2002}
{Kregel} M.,  {van der Kruit} P.~C.,    {de Grijs} R.,  2002, \mnras, 334, 646

\bibitem[\protect\citeauthoryear{{Lucy}}{{Lucy}}{1977}]{lucy1977}
{Lucy} L.~B.,  1977, \aj, 82, 1013

\bibitem[\protect\citeauthoryear{{Macci{\`o}}, {Dutton} \& {van den
  Bosch}}{{Macci{\`o}} et~al.}{2008}]{maccio2008}
{Macci{\`o}} A.~V.,  {Dutton} A.~A.,    {van den Bosch} F.~C.,  2008, \mnras,
  391, 1940

\bibitem[\protect\citeauthoryear{{Makino}, {Sasaki} \& {Suto}}{{Makino}
  et~al.}{1998}]{makino1998}
{Makino} N.,  {Sasaki} S.,    {Suto} Y.,  1998, \apj, 497, 555

\bibitem[\protect\citeauthoryear{{Mihos} \& {Hernquist}}{{Mihos} \&
  {Hernquist}}{1994}]{mihos1994}
{Mihos} J.~C.,  {Hernquist} L.,  1994, \apjl, 425, L13

\bibitem[\protect\citeauthoryear{{Monaghan}}{{Monaghan}}{1992}]{monaghan1992}
{Monaghan} J.~J.,  1992, \araa, 30, 543

\bibitem[\protect\citeauthoryear{{Moore}, {Ghigna}, {Governato}, {Lake},
  {Quinn}, {Stadel} \& {Tozzi}}{{Moore} et~al.}{1999}]{moore1999}
{Moore} B.,  {Ghigna} S.,  {Governato} F.,  {Lake} G.,  {Quinn} T.,  {Stadel}
  J.,    {Tozzi} P.,  1999, \apjl, 524, L19

\bibitem[\protect\citeauthoryear{{Moster}, {Macci{\`o}}, {Somerville},
  {Johansson} \& {Naab}}{{Moster} et~al.}{2010b}]{moster2010b}
{Moster} B.~P.,  {Macci{\`o}} A.~V.,  {Somerville} R.~S.,  {Johansson} P.~H.,
   {Naab} T.,  2010b, \mnras, 403, 1009

\bibitem[\protect\citeauthoryear{{Moster}, {Maccio'}, {Somerville}, {Naab} \&
  {Cox}}{{Moster} et~al.}{2011}]{moster2011a}
{Moster} B.~P.,  {Maccio'} A.~V.,  {Somerville} R.~S.,  {Naab} T.,    {Cox}
  T.~J.,  2011, arXiv:1104.0246

\bibitem[\protect\citeauthoryear{{Moster}, {Somerville}, {Maulbetsch}, {van den
  Bosch}, {Macci{\`o}}, {Naab} \& {Oser}}{{Moster} et~al.}{2010a}]{moster2010a}
{Moster} B.~P.,  {Somerville} R.~S.,  {Maulbetsch} C.,  {van den Bosch} F.~C.,
  {Macci{\`o}} A.~V.,  {Naab} T.,    {Oser} L.,  2010a, \apj, 710, 903

\bibitem[\protect\citeauthoryear{{Naab} \& {Burkert}}{{Naab} \&
  {Burkert}}{2003}]{naab2003}
{Naab} T.,  {Burkert} A.,  2003, \apj, 597, 893

\bibitem[\protect\citeauthoryear{{Naab}, {Jesseit} \& {Burkert}}{{Naab}
  et~al.}{2006}]{naab2006}
{Naab} T.,  {Jesseit} R.,    {Burkert} A.,  2006, \mnras, 372, 839

\bibitem[\protect\citeauthoryear{{Navarro}, {Frenk} \& {White}}{{Navarro}
  et~al.}{1997}]{navarro1997}
{Navarro} J.~F.,  {Frenk} C.~S.,    {White} S.~D.~M.,  1997, \apj, 490, 493

\bibitem[\protect\citeauthoryear{{Negroponte} \& {White}}{{Negroponte} \&
  {White}}{1983}]{negroponte1983}
{Negroponte} J.,  {White} S.~D.~M.,  1983, \mnras, 205, 1009

\bibitem[\protect\citeauthoryear{{Nordstr{\"o}m}, {Mayor}, {Andersen},
  {Holmberg}, {Pont}, {J{\o}rgensen}, {Olsen}, {Udry} \&
  {Mowlavi}}{{Nordstr{\"o}m} et~al.}{2004}]{nordstrom2004}
{Nordstr{\"o}m} B.,  {Mayor} M.,  {Andersen} J.,  {Holmberg} J.,  {Pont} F.,
  {J{\o}rgensen} B.~R.,  {Olsen} E.~H.,  {Udry} S.,    {Mowlavi} N.,  2004,
  \aap, 418, 989

\bibitem[\protect\citeauthoryear{{Oppenheimer} \& {Dav{\'e}}}{{Oppenheimer} \&
  {Dav{\'e}}}{2006}]{oppenheimer2006}
{Oppenheimer} B.~D.,  {Dav{\'e}} R.,  2006, \mnras, 373, 1265

\bibitem[\protect\citeauthoryear{{Pohlen}, {Dettmar}, {L{\"u}tticke} \&
  {Schwarzkopf}}{{Pohlen} et~al.}{2000}]{pohlen2000}
{Pohlen} M.,  {Dettmar} R.-J.,  {L{\"u}tticke} R.,    {Schwarzkopf} U.,  2000,
  \aaps, 144, 405

\bibitem[\protect\citeauthoryear{{Purcell}, {Kazantzidis} \&
  {Bullock}}{{Purcell} et~al.}{2009}]{purcell2009}
{Purcell} C.~W.,  {Kazantzidis} S.,    {Bullock} J.~S.,  2009, \apj, 694, L98

\bibitem[\protect\citeauthoryear{{Qu}, {Di Matteo}, {Lehnert} \& {van
  Driel}}{{Qu} et~al.}{2011}]{qu2011}
{Qu} Y.,  {Di Matteo} P.,  {Lehnert} M.~D.,    {van Driel} W.,  2011, \aap,
  530, A10+

\bibitem[\protect\citeauthoryear{{Quinn} \& {Goodman}}{{Quinn} \&
  {Goodman}}{1986}]{quinn1986}
{Quinn} P.~J.,  {Goodman} J.,  1986, \apj, 309, 472

\bibitem[\protect\citeauthoryear{{Quinn}, {Hernquist} \& {Fullagar}}{{Quinn}
  et~al.}{1993}]{quinn1993}
{Quinn} P.~J.,  {Hernquist} L.,    {Fullagar} D.~P.,  1993, \apj, 403, 74

\bibitem[\protect\citeauthoryear{{Rasmussen}, {Sommer-Larsen}, {Pedersen},
  {Toft}, {Benson}, {Bower} \& {Grove}}{{Rasmussen}
  et~al.}{2009}]{rasmussen2009}
{Rasmussen} J.,  {Sommer-Larsen} J.,  {Pedersen} K.,  {Toft} S.,  {Benson} A.,
  {Bower} R.~G.,    {Grove} L.~F.,  2009, \apj, 697, 79

\bibitem[\protect\citeauthoryear{{Read}, {Lake}, {Agertz} \&
  {Debattista}}{{Read} et~al.}{2008}]{read2008}
{Read} J.~I.,  {Lake} G.,  {Agertz} O.,    {Debattista} V.~P.,  2008, \mnras,
  389, 1041

\bibitem[\protect\citeauthoryear{{Salpeter}}{{Salpeter}}{1955}]{salpeter1955}
{Salpeter} E.~E.,  1955, \apj, 121, 161

\bibitem[\protect\citeauthoryear{{Schwarzkopf} \& {Dettmar}}{{Schwarzkopf} \&
  {Dettmar}}{2000}]{schwarzkopf2000}
{Schwarzkopf} U.,  {Dettmar} R.-J.,  2000, \aap, 361, 451

\bibitem[\protect\citeauthoryear{{Seabroke} \& {Gilmore}}{{Seabroke} \&
  {Gilmore}}{2007}]{seabroke2007}
{Seabroke} G.~M.,  {Gilmore} G.,  2007, \mnras, 380, 1348

\bibitem[\protect\citeauthoryear{{Shaw} \& {Gilmore}}{{Shaw} \&
  {Gilmore}}{1989}]{shaw1989}
{Shaw} M.~A.,  {Gilmore} G.,  1989, \mnras, 237, 903

\bibitem[\protect\citeauthoryear{{Shaw} \& {Gilmore}}{{Shaw} \&
  {Gilmore}}{1990}]{shaw1990}
{Shaw} M.~A.,  {Gilmore} G.,  1990, \mnras, 242, 59

\bibitem[\protect\citeauthoryear{{Somerville}, {Hopkins}, {Cox}, {Robertson} \&
  {Hernquist}}{{Somerville} et~al.}{2008}]{somerville2008a}
{Somerville} R.~S.,  {Hopkins} P.~F.,  {Cox} T.~J.,  {Robertson} B.~E.,
  {Hernquist} L.,  2008, \mnras, 391, 481

\bibitem[\protect\citeauthoryear{{Sommer-Larsen}}{{Sommer-Larsen}}{2006}]{sommerlarsen2006}
{Sommer-Larsen} J.,  2006, \apjl, 644, L1

\bibitem[\protect\citeauthoryear{{Sommer-Larsen}, {G{\"o}tz} \&
  {Portinari}}{{Sommer-Larsen} et~al.}{2003}]{sommerlarsen2003}
{Sommer-Larsen} J.,  {G{\"o}tz} M.,    {Portinari} L.,  2003, \apj, 596, 47

\bibitem[\protect\citeauthoryear{{Soubiran}, {Bienaym{\'e}} \&
  {Siebert}}{{Soubiran} et~al.}{2003}]{soubiran2003}
{Soubiran} C.,  {Bienaym{\'e}} O.,    {Siebert} A.,  2003, \aap, 398, 141

\bibitem[\protect\citeauthoryear{{Springel}}{{Springel}}{2005}]{springel2005a}
{Springel} V.,  2005, \mnras, 364, 1105

\bibitem[\protect\citeauthoryear{{Springel}, {Di Matteo} \&
  {Hernquist}}{{Springel} et~al.}{2005}]{springel2005b}
{Springel} V.,  {Di Matteo} T.,    {Hernquist} L.,  2005, \mnras, 361, 776

\bibitem[\protect\citeauthoryear{{Springel} \& {Hernquist}}{{Springel} \&
  {Hernquist}}{2002}]{springel2002}
{Springel} V.,  {Hernquist} L.,  2002, \mnras, 333, 649

\bibitem[\protect\citeauthoryear{{Springel} \& {Hernquist}}{{Springel} \&
  {Hernquist}}{2003}]{springel2003}
{Springel} V.,  {Hernquist} L.,  2003, \mnras, 339, 289

\bibitem[\protect\citeauthoryear{{Stewart}, {Bullock}, {Wechsler}, {Maller} \&
  {Zentner}}{{Stewart} et~al.}{2008}]{stewart2008}
{Stewart} K.~R.,  {Bullock} J.~S.,  {Wechsler} R.~H.,  {Maller} A.~H.,
  {Zentner} A.~R.,  2008, \apj, 683, 597

\bibitem[\protect\citeauthoryear{{Stinson}, {Bailin}, {Couchman}, {Wadsley},
  {Shen}, {Brook} \& {Quinn}}{{Stinson} et~al.}{2010}]{stinson2010}
{Stinson} G.,  {Bailin} J.,  {Couchman} H.,  {Wadsley} J.,  {Shen} S.,  {Brook}
  C.,    {Quinn} T.,  2010, \mnras, 408, 812

\bibitem[\protect\citeauthoryear{{Toft}, {Rasmussen}, {Sommer-Larsen} \&
  {Pedersen}}{{Toft} et~al.}{2002}]{toft2002}
{Toft} S.,  {Rasmussen} J.,  {Sommer-Larsen} J.,    {Pedersen} K.,  2002,
  \mnras, 335, 799

\bibitem[\protect\citeauthoryear{{Toomre}}{{Toomre}}{1977}]{toomre1977}
{Toomre} A.,  1977, in {B.~M.~Tinsley \& R.~B.~Larson} ed., Evolution of
  Galaxies and Stellar Populations {Mergers and Some Consequences}.
p.~401

\bibitem[\protect\citeauthoryear{{Toth} \& {Ostriker}}{{Toth} \&
  {Ostriker}}{1992}]{toth1992}
{Toth} G.,  {Ostriker} J.~P.,  1992, \apj, 389, 5

\bibitem[\protect\citeauthoryear{{Velazquez} \& {White}}{{Velazquez} \&
  {White}}{1999}]{velazquez1999}
{Velazquez} H.,  {White} S.~D.~M.,  1999, \mnras, 304, 254

\bibitem[\protect\citeauthoryear{{Villalobos} \& {Helmi}}{{Villalobos} \&
  {Helmi}}{2008}]{villalobos2008}
{Villalobos} {\'A}.,  {Helmi} A.,  2008, \mnras, 391, 1806

\bibitem[\protect\citeauthoryear{{Walker}, {Mihos} \& {Hernquist}}{{Walker}
  et~al.}{1996}]{walker1996}
{Walker} I.~R.,  {Mihos} J.~C.,    {Hernquist} L.,  1996, \apj, 460, 121

\bibitem[\protect\citeauthoryear{{White} \& {Rees}}{{White} \&
  {Rees}}{1978}]{white1978}
{White} S.~D.~M.,  {Rees} M.~J.,  1978, \mnras, 183, 341

\bibitem[\protect\citeauthoryear{{Wielen}}{{Wielen}}{1977}]{wielen1977}
{Wielen} R.,  1977, \aap, 60, 263

\bibitem[\protect\citeauthoryear{{Yoachim} \& {Dalcanton}}{{Yoachim} \&
  {Dalcanton}}{2006}]{yoachim2006}
{Yoachim} P.,  {Dalcanton} J.~J.,  2006, \aj, 131, 226

\bibitem[\protect\citeauthoryear{{Zibetti}, {Charlot} \& {Rix}}{{Zibetti}
  et~al.}{2009}]{zibetti2009}
{Zibetti} S.,  {Charlot} S.,    {Rix} H.-W.,  2009, \mnras, 400, 1181

\end{thebibliography}

\label{lastpage}

\end{document}